\documentclass[journal ]{new-aiaa}
\usepackage[utf8]{inputenc}
\usepackage{textcomp}

\usepackage{multicol}
\usepackage{multirow}
\DeclareUnicodeCharacter{2248}{~}
\usepackage{graphicx}
\usepackage{amsmath}
\usepackage[version=4]{mhchem}
\usepackage{siunitx}
\usepackage{longtable,tabularx}
\usepackage{lscape}
\usepackage{rotating}
\usepackage{xcolor}

\title{Investigation of the Effect of Bars on the Properties of Spiral Galaxies: A Multivariate Statistical Study}
\author[Banerjee et al.]{
Prasenjit Banerjee$^{1}$\footnote{Email id: b.prasenjit1994@gmail.com , Telephone number: (+91)9433977094}
, Tanuka Chattopadhyay$^{2}$\footnote{Email id: tchatappmath@caluniv.ac.in , Telephone number: (+91)9433648255}
, Asis Kumar Chattopadhyay$^{1}$\footnote{Email id: akcstat@gmail.com , Telephone number: (+91)9831985850}
\\
$^{1}$Department of Statistics, University of Calcutta, 35, Ballygunge Circular Road, Kolkata-700019, India\\
$^{2}$Department of Applied Mathematics, University of Calcutta, 92, A.P.C Road, Kolkata-700009, India\\
}

\begin{document}

\maketitle

\hrule

\begin{multicols}{2}

\section*{Article Information}
\hrule
\noindent\emph{Article history:}\\
\\
Received XX YY ZZZZ\\
Received in revised form XX YY ZZZZ\\
Accepted XX YY ZZZZ\\
Available online XX YY ZZZZ\\
\\
\\
\hrule
\noindent\emph{Keywords:}\\
\\
\textcolor{black}{
Galaxies: Fundamental parameters, \\
Galaxies: Spirals, \\
Method: Statistical, \\
Astronomical Databases: Miscellaneous \\
}
\\
\\
\\
\\
\\
\\
\\
\hrule
\columnbreak

\section*{Abstract}
\hrule
\begin{abstract}
\vspace{3mm}
Subjective classification of spiral galaxies is not \textcolor{black}{sufficient} for studying the effect of bars on their physical characteristics. In reality \textcolor{black}{the problem} is to comprehend the complex correlations in a multivariate parametric space. Multivariate tools are the best ones for understanding this complex correlation. In this work an objective classification of a \textcolor{black}{large set (26,089)} of  spiral galaxies was \textcolor{black}{compiled} as a value added galaxy catalogue from \textcolor{black}{ sdss DR 15 virtual data archive}.

\textcolor{black}{Initially for dimensionality reduction}, Independent Component Analysis is \textcolor{black}{performed} to determine a set of Independent Components that are linear combinations of 48 observed features (namely ionised lines, Lick indices, photometric and morphological properties). Subsequently a K-means cluster analysis is \textcolor{black}{carried out on the basis of} the 14 best chosen Independent Components to obtain 12 distinct homogeneous groups \textcolor{black}{of spiral galaxies}. Amongst these, 3 groups are the oldest ones (1.6 Gyr - 5.9 Gyr), while 5 groups fall in the medium aged category (1.4 Gyr - 1.6 Gyr), 2 groups consist of only unbarred spirals, 1 group is the youngest one and the remaining one is an outlier. In many groups there are clear indication of recurrent bar formation phenomena which is consistent with few previous simulation works. \textcolor{black}{In order to study the} robustness  a second \textcolor{black}{method} of clustering by Gaussian Mixture Modeling Method (GMMBC) \textcolor{black}{is applied}.
\end{abstract}
\hrule
\end{multicols}

\section{Introduction}
\

Effect of bars on the properties of spiral galaxies is an interesting topic as several studies have shown that many important properties like star formation activity, color, metallicity etc. may depend on the strength of bar (\cite{vera2016effect}; \cite{athanassoula1983formation}; \cite{buta1996galactic}; \cite{combes1993bars}; \cite{martin1995quantitative}; \cite{ellison2011impact}; \cite{zhou2014star}). \textcolor{black}{Many other} works (\cite{debattista1998dynamical}; \cite{weinberg1985evolution}; \textcolor{black}{\cite{debattista2000constraints}; \cite{athanassoula2003determines}; \cite{erwin2019determines}; \cite{kim2020effect}; \cite{garma2020sdss}}) show by numerical simulations that bars can effectively transport gas from the outskirt towards the central regions of the barred galaxies. Subsequently this gas undergo interaction with the edges of the bar which produces shock waves. This shocked gas looses angular momentum which accentuates the flow of gas towards the central region and thus produces starburst. Some works show that bars can be destroyed by the presence of large central mass (\cite{roberts1979gas}; \cite{norman1996bar}; \cite{sellwood1999formation}; \cite{athanassoula2005can}; \textcolor{black}{\cite{spinoso2016bar}; \cite{barbuy2018chemodynamical}; \cite{guo2020new}; \cite{rosas2020buildup}}). This theory indicates that many non-barred disk galaxies might had  bars in the past and thus presence and absence of bars are nothing but a recurrent phenomenon of galaxy life (\cite{bournaud2002gas}; \cite{berentzen2004regeneration}; \cite{gadotti2006lengths}; \textcolor{black}{\cite{pettitt2018bars}; \cite{katz2018gaia}; \cite{hilmi2020fluctuations}}). Many authors have \textcolor{black}{established} that inflow of gas is an efficient mechanism for triggering active galactic nuclei (AGN) and they form bulges or pseudo bulges (\cite{kormendy2004secular}; \cite{debattista2005kinematic}; \cite{debattista2006secular}; \cite{martinez2006evolution}; \cite{aguerri2009origin}; \textcolor{black}{\cite{de2019clocking}; \cite{fragkoudi2020chemodynamics}; \cite{barbuy2018chemodynamical}}).\\

\textcolor{black}{In order to investigate the} relation between bars and host galaxy colors, different studies \textcolor{black}{inferred} that bars are frequently found in late type spiral galaxies those are bluer and less concentrated systems (\cite{barazza2008bars}; \cite{aguerri2009origin}). On the contrary other studies found an excess of barred galaxies with redder colors from different samples (\cite{masters2010galaxy}; \cite{lintott2011galaxy}; \cite{oh2011bar}; \cite{alonso2013effect}; \cite{alonso2014agn}; \cite{vera2016effect}; \textcolor{black}{\cite{kim2020direct}; \cite{cuomo2019bar}}).\\

\textcolor{black}{In connection with} star formation activity, many \textcolor{black}{researchers} \textcolor{black}{indicate} that presence of bars enhance star formation rate (SFR) (\cite{hawarden1986enhanced}; \cite{devereux1987spatial}; \cite{hummel1990environmental}), while several others show that bars do not guarantee increase in star formation activity (\cite{pompea1990test}; \cite{martinet1997bar}; \cite{chapelon1999starbursts}; \textcolor{black}{\cite{donohoe2019redistribution}; \cite{wang2020suppressed}; \cite{kim2017star}; \cite{newnham2020h}}). Similar controversial results are obtained in case of metallicity also (\cite{vila1992relation}; \cite{martin1994influence}; \cite{ellison2011impact}; \cite{sanchez2014stellar}).\\

The above studies \textcolor{black}{are mostly empirical and} drawn conclusion either from simulation \textcolor{black}{studies} or from a control data sets as a whole, which \textcolor{black}{are} prepared in various ways. \textcolor{black}{As the galaxy properties are inter dependent, the corresponding parameters constitute a multivariate set up and for data analysis multivariate techniques have to be used.} Some authors studied the barred galaxy properties by means of bar properties but that is a subjective classification (\cite{vera2016effect}; \cite{carles2016mass}; \textcolor{black}{\cite{kruk2018galaxy}; \cite{seo2019effects}; \cite{cavanagh2020bars}}) which is related to algorithm \textcolor{black}{that} is trained on label data. The labels arise out of pre-established classifications on very few features, \textcolor{black}{or} in other words the algorithm is trained to find results of human subjectivity (eg. Zooniverse \textcolor{black}{\footnote{\url{https://www.zooniverse.org/}}}).\\

In the above context one is tempted to apply statistical (unsupervised) classification \textcolor{black}{e.g.} a multivariate partitioning analysis to explore the homogeneous groups of galaxies, not \textcolor{black}{only} focusing one or few particular aspects of the physics of galaxies but \textcolor{black}{also} by exploring the cluster structure of the data set (\cite{chattopadhyay2019unsupervised}; \cite{fraix2015multivariate}). One basic tool is Principal Component Analysis (PCA). This is used by various authors (\cite{whitmore2012objective}; \cite{cabanac2002classification}; \cite{chattopadhyay2006objective}; \cite{peth2016beyond}). The main \textcolor{black}{target} is the reduction of the dimensionality ( i.e. number of parameters, here). But use of PC to perform a Clustering (unsupervised) is not recommended since the components with largest eigenvalues are the axes of maximum variance and those are generally not the most discriminative ones to reveal the cluster structure (\cite{chang1983using}).\\

\textcolor{black}{For} unsupervised classifications some attempts have been made by K-means cluster analysis \textcolor{black}{on the basis of all the parameters} ( \cite{ellis2005millennium}; \cite{chattopadhyay-chattopadhyay2007globular}; \cite{chattopadhyay2007statistical}; \cite{chattopadhyay2008globular}; \cite{chattopadhyay2009study}; \cite{babu2009horizontal}; \cite{almeida2010automatic}; \cite{fraix2010structures}; \cite{fraix2012six}; \cite{de2016clustering};\cite{modak2017two}; \cite{modak2020unsupervised}).\\

Partitioning of objects into robust groups \textcolor{black}{will be more prominent} when the features are independent. For this, Independent Component Analysis is used which is applicable to a non-Gaussian data set like the present one \textcolor{black}{for dimensionality reduction}. On the basis of the Independent Components (ICs), which are the linear combinations of various observable properties of the spiral galaxies \textcolor{black}{e.g.} broad-band \textcolor{black}{line fluxes} (magnitudes), slopes (colors), medium band line fluxes (Lick indices), the large data set \textcolor{black}{can be} classified (unsupervised) into various homogeneous groups. Then the groups are studied with the help of other estimated properties of the bars, SFR, metallicity, age along with the observed ones.\\

In this study we have prepared a large data set \textcolor{black}{of spiral galaxies} from sdss DR15 and cross-matched with it the corresponding bar properties retrieved from Zooniverse (Galaxy-Zoo). The work \textcolor{black}{contains} several novelties for unsupervised classification \textcolor{black}{as follows}:
\begin{itemize}
    \item A large data set of spiral galaxies from sdss DR15, cross matched with zooniverse.
    \item A large number of observable features.
    \item A large number of estimated features.
    \item The application of ICA.
    \item The justification of ICA to the fact that the present data set is non Gaussian.
    \item Robustness checking by \textcolor{black}{ a widely applicable} method like Gaussian Mixture Modelling Method (GMMBC).
\end{itemize}

In most of the previous studies \textcolor{black}{authors retrieved data from sdss} and they performed supervised classification. ICA has been used widely for source separation (\cite{pires2006sunyaev}; \cite{pike2017application}; \cite{martins2018independent}; \cite{sheldon2018emission}) and dimensionality reduction (\cite{richardson2016interpreting}; \cite{sarro2018estimates}) but rarely for unsupervised classification (\cite{mu2007unsupervised}; \cite{das2015multivariate}; \cite{modak2017two}; \cite{modak2020unsupervised}).\\

The paper is organised as follows: A brief description of the data set is given in Section \ref{section 2}. The methods are descried in Section \ref{section 3}. The results and discussion are included in Section \ref{section 4}. Finally, Section \ref{section 5} concludes the summary.

\section{Data description}\label{section 2}
\

\textcolor{black}{The present} data is a cross matched collection of galaxy catalogues, used for the study of galaxy formation and evolution. It is based on the sdss Data Release 15 (sdss DR15) \footnote{\url{https://skyserver.sdss.org/dr15/en/tools/search/sql.aspx}}.

\subsection{Data Preparataion}
\

\textcolor{black}{The present value added galaxy catalogue of spiral galaxies has been \textcolor{black}{compiled} in the following manner:}
\\
\begin{enumerate}

\item The Baldwin, Phillips \& Telervich diagram ( \textcolor{black}{\cite{baldwin1981classification}; \cite{veilleux1987spectral}; \cite{kauffmann2003host}; \cite{kewley2006host}; \cite{kewley2013theoretical}; Here after, BPT diagram}) \textcolor{black}{(include reference)}  parameters of the galaxies (viz. H$_\beta$ Flux, H$_\alpha$ Flux, Oiii 5007 Flux, Nii 6548 Flux, etc. ) are collected by joining the catalogues GalSpecLine on SpecObj through the spectroscopic object id. 

\item The Spectral Lick Indices of the galaxies (viz. Lick\_Nad, Lick\_Ca 4227, Lick\_g 4300, Lick\_Fe 4383  etc. ) \textcolor{black}{are retrieved} by joining the catalogues GalSpecIndx on SpecObj through the spectroscopic object id. 

\item The Continuum substracted emission EW parameters of the galaxies (viz. $\sigma$-forbidden, Oii 3729 Reqw, H$_\beta$ Eqw, H$_\alpha$ Reqw etc. ) are collected by joining the catalogues GalSpecLine on SpecObj through the spectroscopic object id. 

\item The Star formation rates and the specific star formation rates of the galaxies are collected by joining the catalogues GalSpecExtra on SpecObj through the spectroscopic object id. 

\item The Photometric properties of the galaxies (viz. u, g, r, i, z, petroR90{\_}r, petroR50{\_}r etc. ) are collected from the Galaxy catalogue. 

\item Parameters such as metallicity, logmass, age etc. are collected by joining the catalogues stellarMassStarformingPort on SpecObj through the spectroscopic object id. 

\item The Velocity dispersion the galaxies are collected by joining the catalogues GalSpecInfo on SpecObj through the spectroscopic object id. 

\item The other parameters of the galaxies such as sersic indices, U magnitude etc. are collected from the nsatlas catalogue.

\item The Spiral properties of the galaxies (viz. p{\_}cw, p{\_}acw, p{\_}edge, p{\_}cs etc. ) are obtatined by joining the catalogues zooSpec on SpecObj through the spectroscopic object id. 
\\
\end{enumerate}

Now based on the celestial coordinates ($RA$,$DEC$) of the galaxies these datasets are cross-matched \footnote{\url{http://cdsxmatch.u-strasbg.fr/}} amongst themselves to obtain our master catalogue. For cross matching we have used the `By position' cross match criteria and the radius is taken as 1 arcsec \footnote{\url{http://cdsxmatch.u-strasbg.fr/xmatch/doc/CDSXMatchDoc.pdf}}. The cross match is done throughout the sky, not on the basis of any cone or healpix cells \footnote{\url{http://adass2010.cfa.harvard.edu/ADASS2010/incl/presentations/O01_2.pdf}}. Thus \textcolor{black}{the present} data set contains \textcolor{black}{only} spiral galaxies of different spectroscopic sub classes (viz, UNDEFINED, AGN, AGN{\_}BROADLINE, BROADLINE, STARBURST, STARBURST{\_}BROADLINE, STARFORMING, STARFORMING{\_}BROADLINE etc.). The number of entries are 26,089. Further these master catalogue is crossed with the Hoyle Bar length Catalogue \footnote{\url{https://data.galaxyzoo.org/}}, containing 3150 galaxies. After cross matching the number of unbarred spiral galaxies are 24,320 and the number of barred spirals are 1769. In the present data set the barred spirals are denoted by `1's and the unbarred are denoted by `0's. \\

We have limited the number of \textcolor{black}{parameters} to keep the computation tractable \textcolor{black}{and to reduce noise}. Redundant properties such as Sersic profile in different bands have been eliminated since they \textcolor{black}{more or less bear the same information and} only a few photometric bands and colors have been selected. Our data set consists of low redshift i.e. z $\leq$ 0.06. We have kept most of the important physical information in our data set, so that it did not greatly impact our analysis based on dimentionality reduction through ICA. Now, our final data set consists of 48 parameters which covered spectroscopy, photometry, chemical composition, morphology, and kinematics. All these attributes are described in Table 1 below and details are \textcolor{black}{available} on the sdss website \footnote{\url{http://skyserver.sdss.org/dr15/en/help/browser/browser.aspx##&&history=shortdescr+Tables+U}}.

\section*{Table 1}

	
\noindent Detail description of the parameters of present data set collected for study.

{\renewcommand\arraystretch{1.5}
\noindent\begin{longtable}[c]{|c|c|c|c|}
\hline
\textbf{Parameters}  & \textbf{Description} & \textbf{Parameters}  & \textbf{Description} \\
\hline
\endfirsthead
Lick\_nad & Stellar absorption line (Lick) index & h\_delta\_eqw & The equivalent width for absorption \\ & lines & & \\ \hline

Lick\_cn2 & '' & h\_gamma\_eqw & '' \\ \hline
Lick\_ca4227 & '' & h\_beta\_eqw & '' \\ \hline
Lick\_g4300 & '' & h\_alpha\_eqw & '' \\ \hline
Lick\_fe4383 & '' & $U$ & Absolute magnitude (log of intensity) \\ \hline
Lick\_ca4455 & '' & $G$ & '' \\ \hline
Lick\_fe4531 & '' & $R$ & '' \\ \hline
Lick\_c4668 & '' & $I$ & '' \\ \hline
Lick\_hb & '' & $Z$ & '' \\ \hline
Lick\_fe5015 & '' & $J$ & '' \\ \hline
Lick\_mgb & '' & $H$ & ''\\ \hline
Lick\_fe5270 & '' & $K$ & '' \\ \hline
Lick\_fe5335 & '' & $u-g$ & Apparent magnitude($u$) minus \\ & & & Apparent magnitude ($g$) \\ \hline
Lick\_fe5406 & '' & $g-r$ & Apparent magnitude($g$) minus \\ & & & Apparent magnitude ($r$) \\ \hline
Lick\_fe5709 & '' & $r-i$ & Apparent magnitude($r$) minus \\ & & & Apparent magnitude ($i$) \\ \hline
Lick\_fe5782 & '' & $i-z$ & Apparent magnitude($i$) minus \\ & & & Apparent magnitude ($z$) \\ \hline
Lick\_hd\_a & '' & $u-z$ & Apparent magnitude($u$) minus \\ & & & Apparent magnitude ($z$) \\ \hline
$D_n$(4000) & Break in the spectrum at 4000{\AA} & $J-H$ & Magnitude($J$) minus Magnitude ($H$)  \\ \hline
sigma\_balmer & Velocity dispersion ($\sigma$ not FWHM) & $H-K$ & Magnitude($H$) minus Magnitude ($K$) \\ & measured simultaneously in all & & \\ & of the Balmer lines in Km s$^{-1}$ & & \\ \hline
sigma\_forbidden & Velocity dispersion ($\sigma$ not FWHM) & \multicolumn{2}{c|}{\textbf{The following parameters have been used for further}} \\ & measured simultaneously in all & \multicolumn{2}{c|}{\textbf{study after finding the homogeneous groups}} \\ & the forbidden lines in Km s$^{-1}$ & \multicolumn{2}{c|}{ }\\ \hline
oii\_3729\_reqw & The equivalent width of the continuum- & Metallicity & Metallicity of best fit template \\ & subtracted emission line with the & & (5 categories : 0.004, 0.01, \\ & other emission lines subtracted off & & 0.02, 0.04, or "composite") \\ & (EW\_stellar = REQW - EQW) & & \\ \hline
neiii\_3869\_reqw & '' & $z$ & Redshift \\ \hline
oiii\_5007\_reqw & '' & SFR ($M_{\odot} yr^{-1}$)& Star-formation rate of best fit \\ \hline
hei\_5876\_reqw & '' & log($M_*$) ($M_\odot$) & Best-fit stellar mass of galaxy \\ \hline
oi\_6300\_reqw & '' & Age ($Gyr$)& Age of best fit \\ \hline
h\_alpha\_reqw & '' & un(0)barred(1) & Indicator variable for barred(1) \\ & & & and unbarred(0) galaxies\\ \hline
nii\_6584\_reqw & '' & \multicolumn{2}{c|}{\textbf{For barred galaxies}}\\ \hline
sii\_6731\_reqw & '' & length\_scaled & The ratio of bar length to galaxy size \\ \hline
c & Concentration index: & length\_avg & The average bar length scatter\\ & Sersic\_r90\_R  /  Sersic\_r50\_R & & per observer, averaged \\ & & & over galaxies being observed \\ \hline

\end{longtable}}

The initial 48 parameters in the above table are used for the data analysis and the rest of the parameters, mentioned above, have been used for further study.

\begin{figure}[htb!]
\centering
\includegraphics[width=0.75\textwidth]{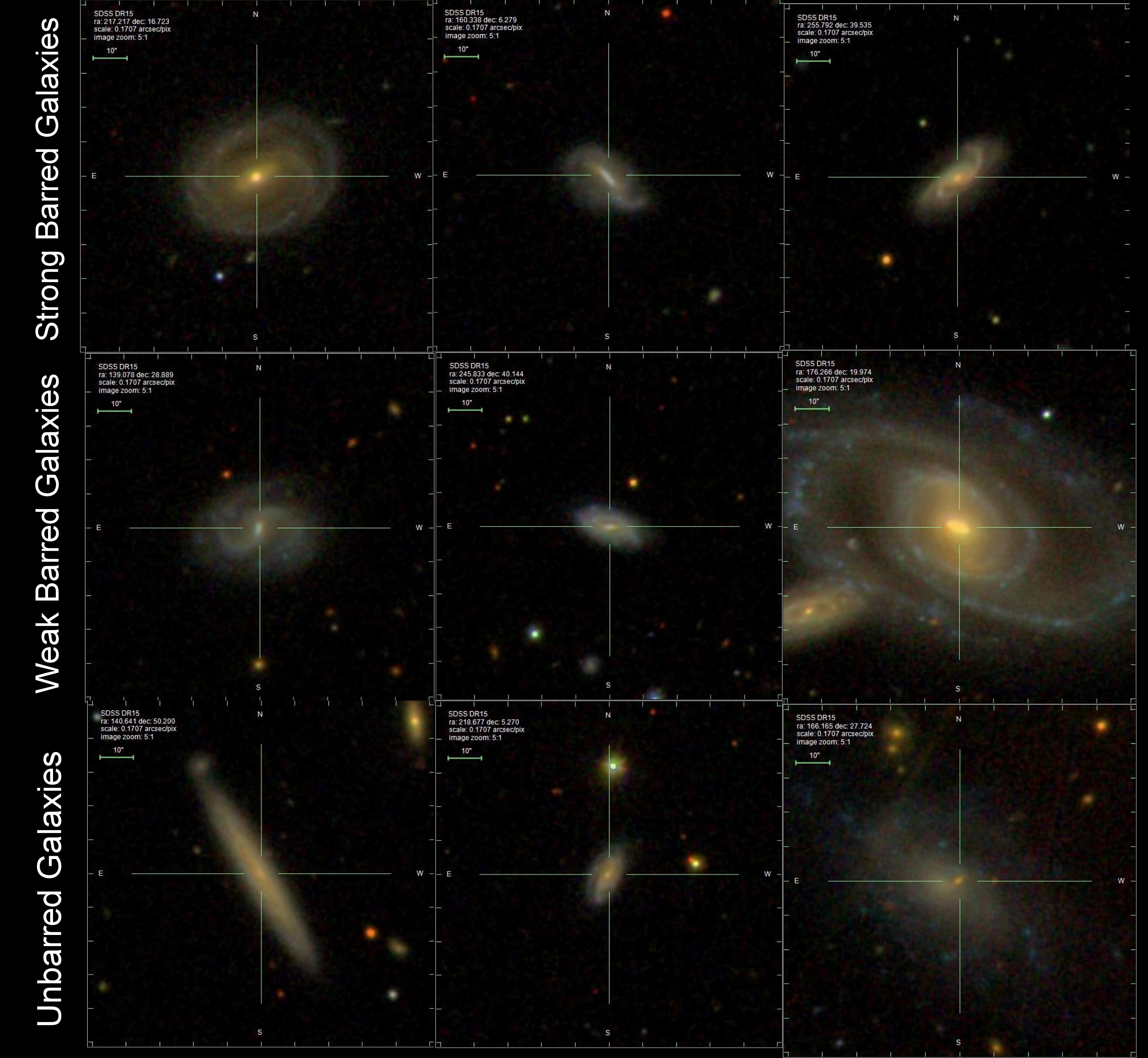}
\caption*{Fig. 1: Images of typical examples of galaxies used in our data set classified as strong barred, weak barred and unbarred galaxies.}
\end{figure}

\section{Statistical Methods}\label{section 3}
\

Thus, we are left with a data set containing 26,089 \textcolor{black}{spiral} galaxies with 48 different parameters \textcolor{black}{ for statistical analysis}. The data set is quite large for computational tractability. \textcolor{black}{Therefore} we use the \textcolor{black}{dimensionality reduction (ICA)} technique to reduce the size of the data without loosing any vital information from it and \textcolor{black}{subsequently} we have clustered the data set to observe any coherent groups present in it.

\subsection{Test for Gaussianity}
\

There are several way for testing the Gaussianity of the data set, such as Shapiro-Wilk test (\cite{shapiro1965analysis}), Anderson Darling Test (\cite{stephens1974edf}), Kolmogorov-Smirnoff test (\cite{lilliefors1967kolmogorov}) etc. Here we have used the Shapiro-Wilk Test for checking the Gausianity of our data set, any other method could have easily opted for. In this technique the test statistics is defined as
\begin{center}
    \centering {W = $\frac{\sum_{i=1}^{n}a_ix_i^2}{\sum_{i=1}^{n}(x_i-\Bar{x})^2}$},
\end{center}
$n$ being the number of observations, $x_i$'s are the ordered sample values and $a_i$'s are the constants generated from the order statistics of a sample from a normal distribution. As our data set is multivariate in nature, a multivariate extension of the Shapiro-Wilk test (\cite{villasenor2009generalization}) have been used. The p-value of the test comes out to be $2.2\times10^{-16}$, which is quite small. Hence we are more inclined in rejecting the null hypothesis, that our data is from a Gaussian setup i.e. the data set is found out to be Non-Gaussian in nature.

\subsection{Independent Component Anlysis}
\

PCA has been applied by many authors (\cite{brosche1973manifold}; \cite{whitmore2012objective}; \cite{murtagh2012multivariate}, etc) \textcolor{black}{for several purposes} but it is not appropriate for clustering and classification (\cite{chang1983using}). Moreover, one of the \textcolor{black}{inherent feature} in PCA is \textcolor{black}{that}, the data set should be a Gaussian data, but our data set is a non-Gaussian one. ICA is a dimension reduction technique, i.e., it reduces the number of observed parameters p to a pre-defined number m (where, m $\ll$ p) of new variables (\textcolor{black}{here the significant IC components)}. This technique is mainly applicable to non-Gaussian setup (\cite{hyvarinen1998independent}; \cite{hyvarinen1999afixed}; \cite{hyvarinen1999bgaussian}; \cite{pfister2019robustifying}). Another basic difference between ICA and PCA is that, in PCA the components are assumed to be uncorrelated but not independent where as in case of ICA the components are assumed to be mutually independent amongst each other. For further details regarding the comparison between these two \textcolor{black}{one can consult Section 3 of \cite{chattopadhyay2013aindependent}, and references therein.}

\subsubsection{Method: Independent Component Analysis}\label{section 3.2.1}
\

let $X_1, X_2, X_3, ..., X_p$ be $p$ random vectors (here, $p$ = 48) and $n$ (here, $n$ = 26,089) be the number of observations for each $X_i$, (i = 1, 2, 3, ..., p).\\

Let $X=AS$, where, $S$=$\{S_1, S_2, S_3, ..., S_p\}^\prime$ is a random vector of hidden components $S_i$'s, ($i$ = 1, 2, 3, ..., $p$). $A$ is a non-singular matrix, also known as the mixing matrix. $S_i$'s are mutually independent amongst themselves. The objective of ICA is to find $S$ by inverting $A$, i.e., $S$ = $A^{-1}X$ = $WX$, where, $A^{-1} = W$. $W$ is called the unmixing matrix as it is the inverse of $A$, the mixing matrix. ICA separates the Independent Components (ICs) (sources) present in a \textcolor{black}{mixture} (\cite{comon1994independent}; \cite{chattopadhyay2013bindependent}). To obtain independence, the non-Gaussianity of the data is maximised using negentropy. There are several techniques for ICA such as FastICA, ProDenICA (\cite{hastie2003independent}), KernelICA etc. One of them is FastICA algorithm (\cite{hyvarinen2000independent}). In this method the ICs are estimated one by one. This algorithm converges very fast and is very reliable. It is the most commonly used algorithm and is also very easy to use.\\

There is no good method available for the determination of the optimum number of ICs. We generally choose it by using the optimum number of PCs (irrespective of the data being non-Gaussian) (\cite{albazzaz2004statistical}; \cite{chattopadhyay2013bindependent}; \cite{eloyan2013semiparametric}), to find m (m$\ll$p) (\cite{chattopadhyay-chattopadhyay2007globular}; \cite{babu2009horizontal}; \cite{fraix2010structures}; \cite{chattopadhyay2010statistical}; \cite{chattopadhyay2013bindependent}). Another novel criterion of choosing the optimal number of ICs is Maximally Stable Transcriptome Dimension (MSTD) (\cite{kairov2017determining}). \\

This technique depends on a fundamental parameter M (effective dimension of the data, here 48), as well as the number of ICs computed) whose effects are being investigated on the stability of the ICs. The range of M values are from $M_{min}$ (here, 2) to $M_{max}$ (here, 40). For each M ranging from $M_{min}$ to $M_{max}$, the data dimension is being reduced to M by PCA and then data has been whitened. Afterwards, in the whitened space the actual signal decomposition is applied by defining M new axes. Each of them maximize the non-Gaussianity of data point projections distribution \textcolor{black}{(Fig. 2).}

\subsubsection*{The algorithm for determining the MSTD:}
\
\begin{enumerate}

\item Define two numbers $M_{min}$ and $M_{max}$ \textcolor{black}{which} denote the maximal and minimal possible numbers of computed \textcolor{black}{independent} components, respectively.\\

\item Define a number $T$ (here, $T$=100). It denotes the number of ICA runs for estimating the components stability.\\

\item For each $M$ ranging between $M_{min}$ and $M_{max}$ :\\

\begin{enumerate}[label=3.\alph*)]
 
\item Find out the $M$, ICs using the fastICA algorithm and iterate it for $T$ times. Thus we will get a data set of $ M\times T$, ICs.\\

\item Now, cluster the newly formed $M\times T$, components into $M$ clusters using the agglomerative hierarchical clustering algorithm, where the measure of dissimilarity being 1- |$r_{ij}$|. $r_{ij}$ is the Pearson's correlation coefficient between the components.\\

\item For each cluster $C_k$ out of $M$ clusters ($C_1$, $C_2$,…, $C_M$), the stability index is obtained, using the formula mentioned below:\label{3.3.c}

\begin{center}
    \centering {$I_q(C_k)=\frac{1}{|C_k|^{2}}\sum\limits_{i,j\in C_k}|r_{ij}|$}-{$\frac{1}{|C_k|\sum\limits_{l\neq k}|C_l|}\sum\limits_{i\in C_k}\sum\limits_{j\notin C_k}|r_{ij}|$}
\end{center}

\vspace{1mm}
where, |$C_k$| denotes the size of the $k^{th}$ cluster.\\

\item The Average Stability Index (ASI) for M clusters is now given by:

\begin{center}
    \centering {S(M)=$\frac{1}{M}\sum\limits_{k}I_q(C_k)$}
\end{center}

\end{enumerate}
\vspace{1mm}

\item 
MSTD is given as the point of intersection of the two lines (\textcolor{black}{magenta} and the green lines in Fig. 3) approximating the distribution of stability profiles. The lines are computed using a simple k-lines clustering algorithm (\cite{agarwal2005beyond}) for k = 2. Here MSTD comes out to be 14 (black line in Fig. 2).

\end{enumerate}
\vspace{3mm}
The clustering quality index as mentioned under \ref{3.3.c} is used here. It measures the quality of the clustering of ICs after multiple runs with random initial conditions by taking the difference between the average intra-cluster similarity and the average inter-cluster similarity \textcolor{black}{(\cite{himberg2004validating}).}
\\

It is further hypothesized that the point of inflection in the distribution of the stability profiles indicates the optimal number of ICs (Fig. 2). To find that point, the stability measures are clustered along the two lines (similar to 2-means clustering, but here the lines are taken as centroids instead of points) (\textcolor{black}{\cite{feldman2003algorithms}}). In this technique, the line (Fig. 2, red line) with positive slope grouped the stability profiles with lower values of M, while another line (Fig. 2, blue line) matched the stability components for the rest. This intersection of these lines (Fig. 2, black line) provided a consistent estimate of the effective number of ICs. This estimate is known as Maximally Stable Transcriptome Dimension (MSTD). This estimate is free of parameters (thresholds) unlike various information theory based criteria (BIC, AIC). It exploits the qualitative change in the character of the stability profile in higher dimensional data.\\

\begin{figure}[htb!]
\centering
\includegraphics[width=0.75\textwidth]{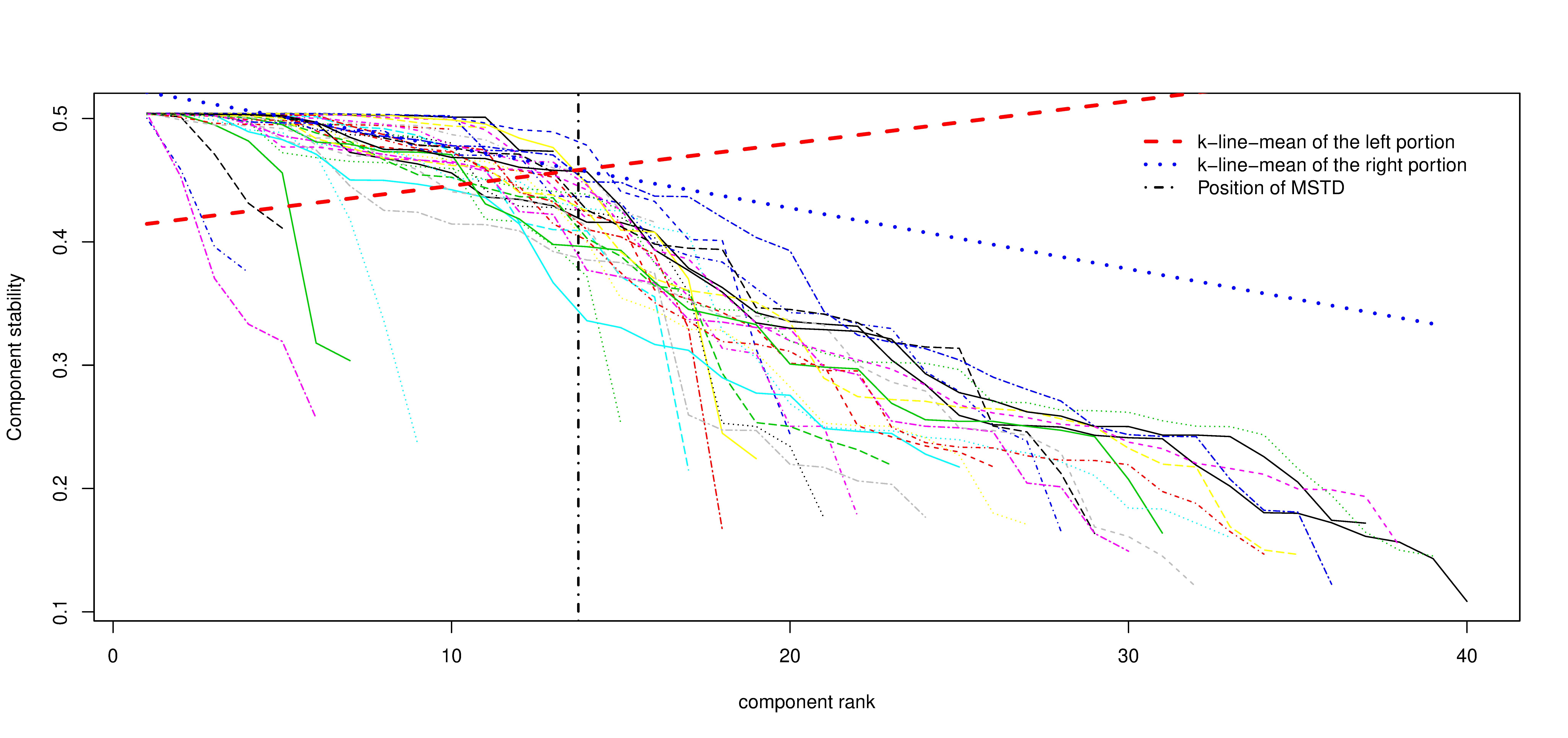}
\caption*{Fig. 2: Stability Index for ICA decompositions in various
dimensions (from 2 to 40).}
\end{figure}

\begin{figure}[htb!]
\centering
\includegraphics[width=0.75\textwidth]{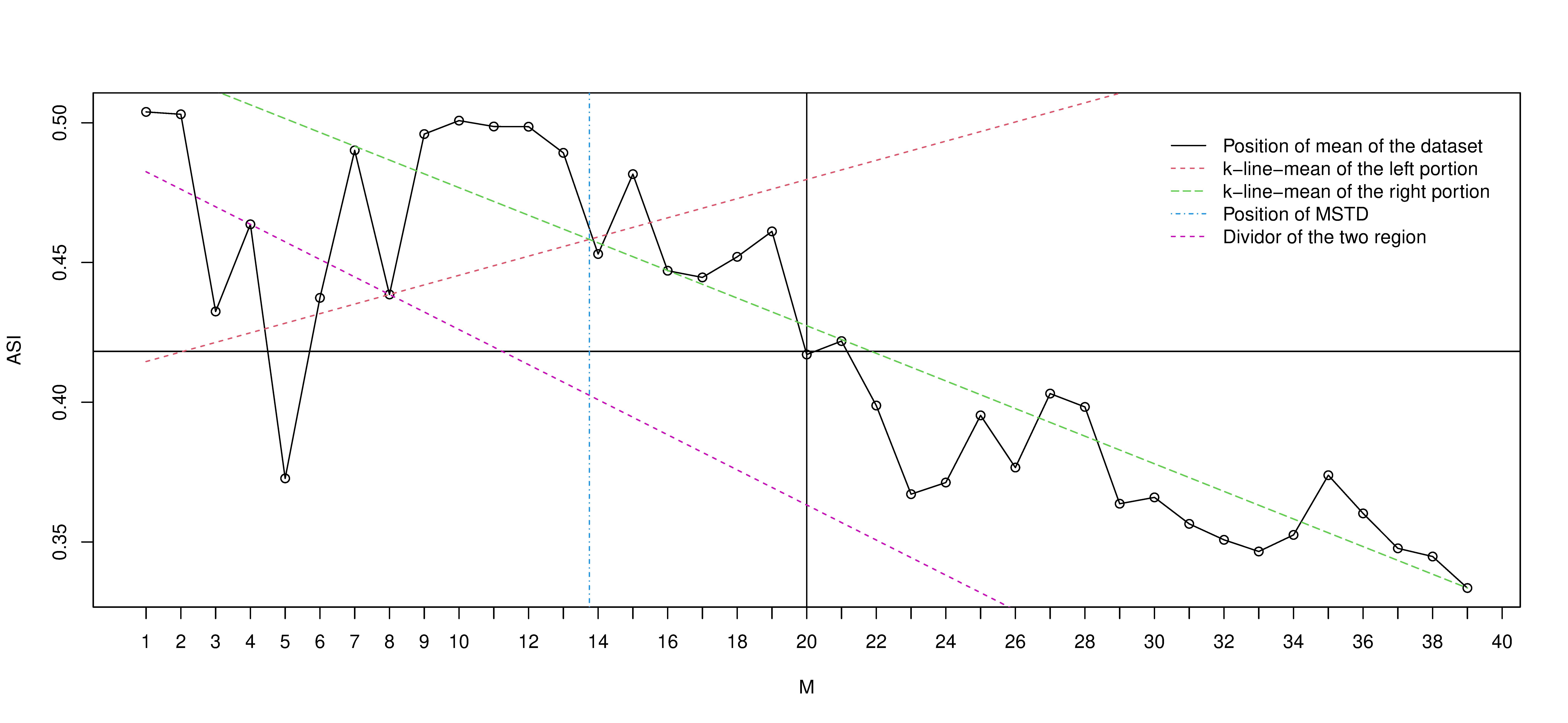}
\caption*{Fig. 3: Average Stability Index along with the two-line clustering result is shown by green and red dashed lines, with MSTD determined as the point of their intersection (vertical turquoise line).}
\end{figure}


Three major conclusions can be made from the figures:\\

\begin{enumerate}

\item With the increase of M, the average stability of the computed components ($S_M^{Total}$) decreases (Fig. 3).\\

\item $S_M^{Total}$ is characterized by the presence of local maxima, defining certain distinguished values of M that correspond to the (locally) maximally stable set of components (Fig. 3).\\

\item The stability profiles for various values of M can be classified into two, viz. (a) Stability values for which the value of M is low (M upto 9), \textcolor{black}{in this case the stability values are unstable, it shows an irregularity in the trend of the stability profiles over M} and (b) the stability values for those forming a large proportion of the components with higher values of M, \textcolor{black}{in this case the stability values are stable and gradually decreasing with M} (Fig. 2).\\

\end{enumerate}

\subsection{Cluster Analysis}
\

In our data set we have applied the K-means clustering technique and its robustness in further examined by another clustering technique, viz. Gaussian Mixture Model Based Clustering (GMMBC). In this work we have done clustering on the basis of the 14 Independent Components, which includes spectroscopic, photometric, chemical composition, morphological, and kinematic properties of galaxies.

\subsubsection{Optimal choice of clusters}
\

There are several methods for finding the optimal number of clusters, present in the data e.g. technique by \cite{sugar2003finding}, gap statistics (\cite{tibshirani2001estimating}) and many more . If an inherent clustering is present in the data then it is manifested by any clustering technique. We have used the Dunn index to find out the optimal number of cluster (\cite{dunn1974well}) under the method of K-means clustering. Dunn index takes value between 0 to $\infty$. Initially we  determined the structures of sub populations (clusters) for varying number of clusters say, k = 1, 2, 3, 4, ......etc. For each such clusters, we have computed the values of the Dunn Index. The value for which the Dunn Index comes out to be highest (here, 0.0063) is taken as the optimal choice of cluster (here, 12) present in our data set and that is the optimal choice of k taken in K-means analysis (Fig. 4).
This finding is further justified by another technique i.e, GMMBC.

\begin{figure}[htb!]
\centering
\includegraphics[width=0.5\textwidth]{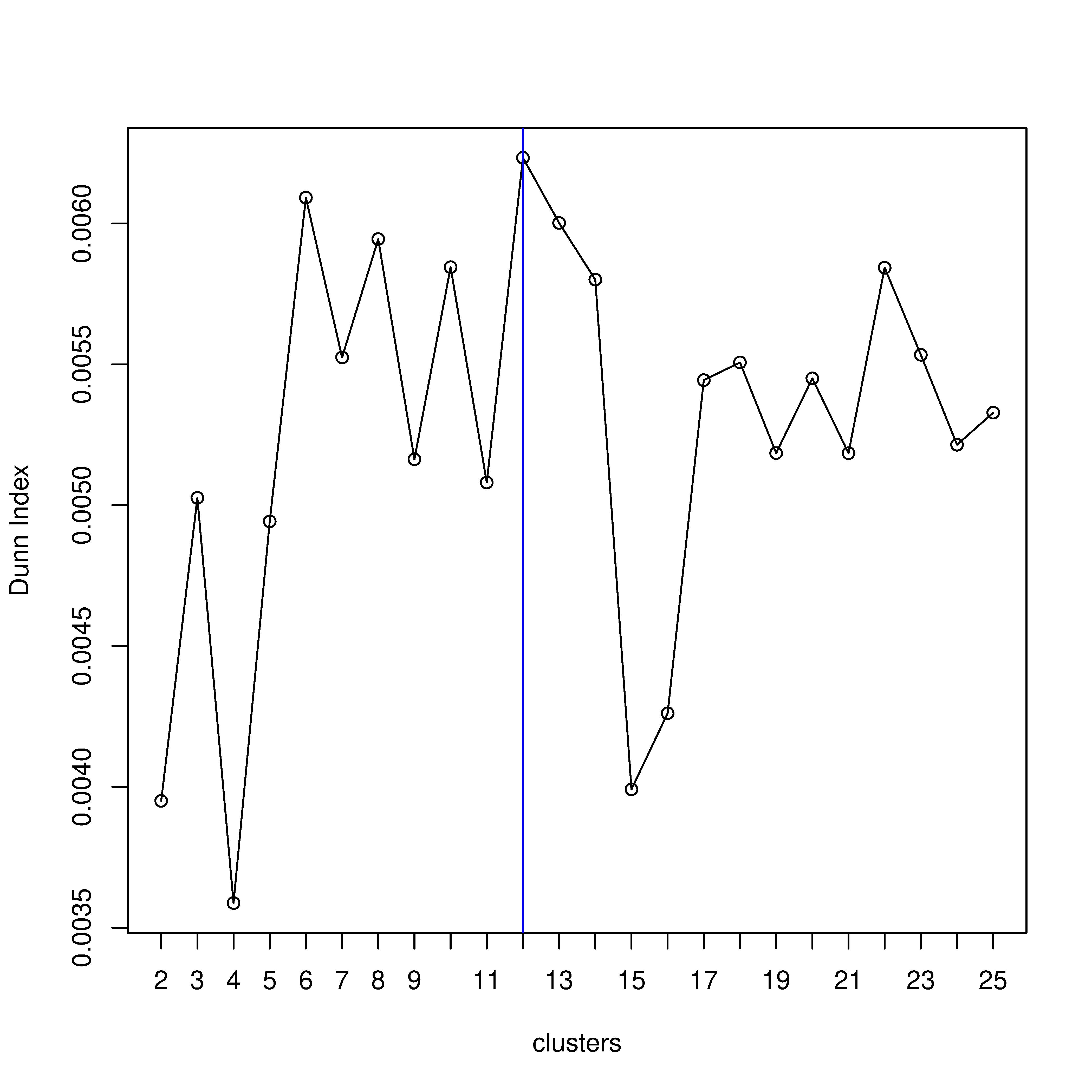}
\caption*{Fig. 4: Dunn Index for different k.}
\end{figure}

In this technique also, we have initially varied k over a certain range (say 1, 2, 3, ....etc.). As GMMBC is a model based clustering, the choices of k indicate the choice of the number of models being mixed among themselves and that mixture of models is taken for which the value of BIC is lowest. Here also the optimal number of cluster (i.e, the optimal number of models) appears to be 12 with a BIC value of -2.9$\times10^{5}$ (Fig. 5).

\begin{figure}[htb!]
\centering
\includegraphics[width=0.75\textwidth]{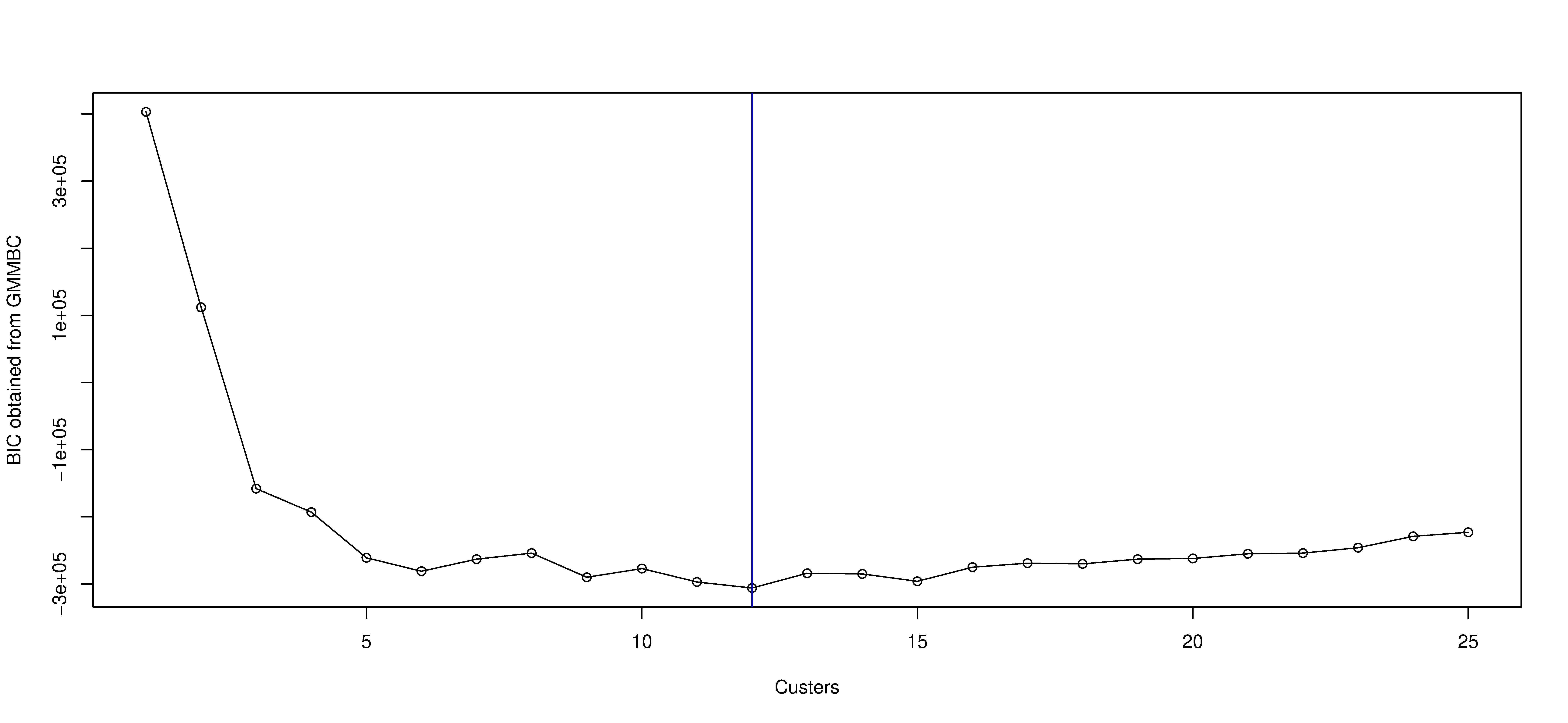}
\caption*{Fig. 5: BIC values obtained from GMMBC for various number of clusters.}
\end{figure}

\subsubsection{Final selection of the reduced data set}\label{section 3.3.2}
\

Thus we have determined that the data set contains 12 distinct groups amongst themselves and from the previous section, the dimension of the data set can be reduced to 14 from 48 set of parameters using MSTD technique (Section \ref{section 3}). We have used two different indexes for choosing the best set of 14 ICs from the complete set of 48 ICs \textcolor{black}{namely} Average Silhouette Width (ASW) (\cite{rousseeuw1987silhouettes}) and Within cluster Sum of Squares (WSS). WSS is used to check whether a clustering is good or poor. It is the sum of squares of the distances of the observations (x) within a cluster ($C_i$) from the cluster centroid ($r_i$). It is given as:

\begin{center}
    \centering {$WSS=\sum\limits_{x\in C_i}(x-r_i)^2$}
\end{center}

A good clustering yields a small within cluster sum of squares. In this way we choose that particular set of \textcolor{black}{14 }ICs which gives the best clustering with respect to both ASW as well as WSS.\\

In order to choose 14 ICs at random out of 48 under the above mentioned method we have used systematic sampling. Systematic sampling  \textcolor{black}{(\cite{rao1988estimation})} is a type of probability sampling method, in which the samples are drawn from a larger population according to a random starting point but with a fixed and periodic interval. This interval is called the sampling interval. We have used a systematic sampling scheme for the choice of 14 ICs from the set of 48 ICs. We further extend this sampling technique to all possible combinations of the initial random start so that for any random start we have a set of sample in our hand. Moreover we have taken all possible combinations of the sampling intervals also so that for any choice of the sampling interval we have a corresponding set of sample in our hand.\\

In doing this, some of the samples gets repeated, such as if we start with IC number 1 and goes on taking those ICs with a periodic sampling interval of 4, (with a \textcolor{black}{maximum} periodicity of 48, as we have 48 ICs in our hand) then we end up taking IC numbers 1,5,9,13,17,21,25,29,33,37,41,45,1,5 respectively. Here IC number 1 \& 5 are repeated in the set, hence this set of ICs are not taken in our analysis. In this way those samples are rejected for our analysis where the ICs gets repeated in choosing the set of 14 different ICs from a sample of 48 ICs.\\

After getting all the set of ICs we calculate the ASW and WSS (mentioned earlier) for each samples and took that particular sample as our reduced data which gives the maximum ASW or the minimum WSS. ICs starting from number 39 and with a sample interval of 23 (viz., IC39, IC14, IC37, IC12, IC35, IC10, IC33, IC8, IC31, IC6, IC29, IC4, IC27, IC2) gives an ASW of 0.188904 \& WSS of 5.76606 and satisfies our aforementioned criteria for the selection of the sample. We further rename the set of ICs  (IC39, IC14, IC37, IC12, IC35, IC10, IC33, IC8, IC31, IC6, IC29, IC4, IC27, IC2) as \textcolor{black}{(IC1 - IC14)} respectively for the sake of our simplicity in addressing the ICs.\\

Finally, we have applied K-means \textcolor{black}{cluster analysis} on this data set by using 14 optimally selected Independent Components as parameters and the value of k as 12. Further, a second clustering by GMMBC is performed over \textcolor{black}{the same} data set for robustness analysis (Table 2).

\section*{Table 2}
\noindent(The median values of the Age (Gyr), length\_avg along with the standard error ( in parentheses ), and the mean values of the observable parameter along with the standard error ( in parentheses ), for each cluster is given below. The clusters obtained by K-means are denoted by `K' and the clusters obtained by GMMBC are denoted by `G'. \textcolor{black}{The mean values for rest of the observable parameter along with the standard error ( in parentheses ) is given in the following Url: \url{https://drive.google.com/file/d/1BnELad32A6VyDahR4txNpVBWGxvrN8SE/view?usp=sharing}})

\begin{longtable}[c]{ | c | c | c | c | c |}
\hline
	\textbf{Cluster Indices} & \textbf{Sample\_Size} & \textbf{Age (Median Values)} & \textbf{Barred\_sample\_Size} & \textbf{length\_avg (Median Values)} \\ \hline
	\textbf{\emph{K1}} & 1616 & 2.3(0.05424) & 301 & 8.87097(0.21045) \\ \hline
	\textbf{\emph{K2}} & 5257 & 1.609(0.03186) & 399 & 9.2719(0.17604) \\ \hline
	\textbf{\emph{K3}} & 4103 & 1.609(0.03445) & 143 & 8.05755(0.26768) \\ \hline
	\textbf{\emph{K4}} & 126 & 5.875(0.3058) & 15 & 10.68616(0.8906) \\ \hline
	\textbf{\emph{K5}} & 425 & 1.0152(0.11685) & 12 & 7.07733(0.63335) \\ \hline
	\textbf{\emph{K6}} & 17 & 0.2273(1.55877) & 1 & 10.34275(0) \\ \hline
	\textbf{\emph{K7}} & 23 & 0.7187(0.80196) & 1 & 8.10091(0) \\ \hline
	\textbf{\emph{K8}} & 3723 & 1.434(0.03455) & 150 & 7.40862(0.26954) \\ \hline
	\textbf{\emph{K9}} & 1 & 13.25(0) & 0 & NA \\ \hline
	\textbf{\emph{K10}} & 185 & 1.434(0.14679) & 4 & 5.36336(0.40655) \\ \hline
	\textbf{\emph{K11}} & 6365 & 2.1(0.03878) & 570 & 10.13034(0.16852) \\ \hline
	\textbf{\emph{K12}} & 4248 & 1.434(0.03203) & 173 & 8.70386(0.27095) \\ \hline \hline
	\textbf{Cluster Indices} & \textbf{Sample\_Size} & \textbf{Age (Median Values)} & \textbf{Barred\_sample\_Size} & \textbf{length\_avg (Median Values)} \\ \hline
	\textbf{\emph{G1}} & 432 & 1.434(0.08072) & 8 & 7.06626(0.97356) \\ \hline
	\textbf{\emph{G2}} & 3088 & 1.8(0.03348) & 363 & 8.77866(0.20282) \\ \hline
	\textbf{\emph{G3}} & 6803 & 1.2781(0.02093) & 178 & 6.78002(0.21122) \\ \hline
	\textbf{\emph{G4}} & 10697 & 1.9(0.02655) & 816 & 9.75887(0.127) \\ \hline
	\textbf{\emph{G5}} & 101 & 2.75(0.5013) & 7 & 10.03251(1.11173) \\ \hline
	\textbf{\emph{G6}} & 759 & 1.8(0.12186) & 13 & 10.61704(1.21821) \\ \hline
	\textbf{\emph{G7}} & 2004 & 1.434(0.04803) & 43 & 8.27783(0.37798) \\ \hline
	\textbf{\emph{G8}} & 704 & 1.0152(0.07775) & 23 & 7.33386(0.60604) \\ \hline
	\textbf{\emph{G9}} & 811 & 2.4(0.10432) & 190 & 10.01833(0.28531) \\ \hline
	\textbf{\emph{G10}} & 416 & 5.5(0.17722) & 76 & 10.26973(0.46843) \\ \hline
	\textbf{\emph{G11}} & 270 & 1.68(0.17767) & 51 & 8.14877(0.5025) \\ \hline
	\textbf{\emph{G12}} & 4 & 10(2.61632) & 1 & 11.65327(0) \\ \hline
\end{longtable}

\begin{landscape}
\begin{longtable}[c]{ | c | c | c | c | c | c | c | c |}
\hline
	\textbf{Cluster Indices} & \textbf{Sample\_Size} & \textbf{$log(M_*)$} & \textbf{$D_n$(4000)} & \textbf{U} & \textbf{G} & \textbf{R} & \textbf{I} \\ \hline
	\textbf{\emph{K1}} & 1616 & 9.90439(0.00835) & 1.2544(0.00299) & -19.43732(0.01757) & -20.70913(0.01635) & -21.3206(0.01633) & -21.67598(0.01666) \\ \hline
	\textbf{\emph{K2}} & 5257 & 9.86208(0.00661) & 1.46626(0.00287) & -18.8964(0.01304) & -20.41194(0.01188) & -21.0993(0.01197) & -21.47548(0.01205) \\ \hline
	\textbf{\emph{K3}} & 4103 & 9.72752(0.00699) & 1.42718(0.00278) & -18.69309(0.01346) & -20.17005(0.0125) & -20.84024(0.01263) & -21.21494(0.01267) \\ \hline
	\textbf{\emph{K4}} & 126 & 10.72587(0.03902) & 1.90489(0.01165) & -19.19173(0.07196) & -21.11759(0.07008) & -21.98127(0.07163) & -22.40291(0.07261) \\ \hline
	\textbf{\emph{K5}} & 425 & 9.1308(0.02185) & 1.18082(0.00562) & -18.61763(0.03978) & -19.72759(0.04088) & -20.14142(0.04274) & -20.40716(0.0445) \\ \hline
	\textbf{\emph{K6}} & 17 & 7.83294(0.37162) & 1.40902(0.04798) & -17.14303(0.61724) & -17.79457(0.73786) & -21.02199(0.33941) & -21.18044(0.33999) \\ \hline
	\textbf{\emph{K7}} & 23 & 8.79826(0.06862) & 1.04913(0.00631) & -18.33515(0.14112) & -19.28366(0.12926) & -19.60257(0.11795) & -19.78258(0.12601) \\ \hline
	\textbf{\emph{K8}} & 3723 & 9.62969(0.00748) & 1.39598(0.00307) & -18.64067(0.01387) & -20.06829(0.01294) & -20.7063(0.01308) & -21.06136(0.01333) \\ \hline
	\textbf{\emph{K9}} & 1 & 8.01(0) & 1.28818(0) & -9.37978(0) & -17.25025(0) & -17.85383(0) & -18.1867(0) \\ \hline
	\textbf{\emph{K10}} & 185 & 9.46335(0.03027) & 1.35396(0.01373) & -18.30381(0.06301) & -19.70998(0.05604) & -20.33062(0.05465) & -20.67803(0.05457) \\ \hline
	\textbf{\emph{K11}} & 6365 & 10.1918(0.00699) & 1.57024(0.003) & -19.15221(0.01021) & -20.76579(0.01048) & -21.5271(0.011) & -21.9312(0.01143) \\ \hline
	\textbf{\emph{K12}} & 4248 & 9.72668(0.00698) & 1.42778(0.00278) & -18.73717(0.01316) & -20.2036(0.01229) & -20.86652(0.01239) & -21.23347(0.01264) \\ \hline \hline
	\textbf{Cluster Indices} & \textbf{Sample\_Size} & \textbf{$log(M_*)$} & \textbf{$D_n$(4000)} & \textbf{U} & \textbf{G} & \textbf{R} & \textbf{I} \\ \hline
	\textbf{\emph{G1}} & 432 & 9.365(0.01722) & 1.34264(0.00994) & -18.10025(0.03795) & -19.50202(0.03409) & -20.12297(0.03315) & -20.476(0.0332) \\ \hline
	\textbf{\emph{G2}} & 3088 & 9.78574(0.00719) & 1.29102(0.00256) & -19.22077(0.01415) & -20.52803(0.01358) & -21.13111(0.0138) & -21.47963(0.01412) \\ \hline
	\textbf{\emph{G3}} & 6803 & 9.55841(0.00466) & 1.35675(0.00167) & -18.65592(0.00974) & -20.04478(0.00886) & -20.65707(0.00872) & -21.00623(0.00881) \\ \hline
	\textbf{\emph{G4}} & 10697 & 10.12099(0.00453) & 1.57668(0.00195) & -19.06244(0.00805) & -20.67267(0.00772) & -21.42848(0.0078) & -21.83013(0.00788) \\ \hline
	\textbf{\emph{G5}} & 101 & 9.33158(0.13348) & 1.39201(0.01821) & -18.20581(0.18508) & -19.51398(0.21045) & -20.65595(0.17291) & -20.92904(0.18339) \\ \hline
	\textbf{\emph{G6}} & 759 & 9.84625(0.01795) & 1.51084(0.00884) & -18.14955(0.04125) & -19.90522(0.03106) & -20.68149(0.0311) & -21.106(0.03147) \\ \hline
	\textbf{\emph{G7}} & 2004 & 9.63086(0.0095) & 1.41727(0.00349) & -18.45827(0.01884) & -19.94553(0.01697) & -20.62005(0.01678) & -20.99567(0.01688) \\ \hline
	\textbf{\emph{G8}} & 704 & 9.12939(0.01529) & 1.20886(0.0035) & -18.48432(0.02834) & -19.64376(0.02843) & -20.08478(0.02932) & -20.35706(0.03059) \\ \hline
	\textbf{\emph{G9}} & 811 & 10.26343(0.01468) & 1.39753(0.0079) & -19.53831(0.02447) & -21.04197(0.02272) & -21.78353(0.02314) & -22.18099(0.02503) \\ \hline
	\textbf{\emph{G10}} & 416 & 10.77909(0.01898) & 1.88879(0.00606) & -19.40894(0.03545) & -21.30414(0.03478) & -22.17206(0.03502) & -22.60102(0.03539) \\ \hline
	\textbf{\emph{G11}} & 270 & 9.59041(0.03426) & 1.19806(0.00966) & -19.23462(0.05048) & -20.40093(0.05429) & -20.90591(0.05899) & -21.21069(0.06243) \\ \hline
	\textbf{\emph{G12}} & 4 & 8.86(0.68627) & 1.51563(0.11345) & -13.46851(2.36757) & -16.72492(2.0779) & -17.79778(1.81164) & -18.25704(1.74008) \\ \hline

\end{longtable}
\end{landscape}

\begin{landscape}
\begin{longtable}[c]{ | c | c | c | c | c | c | c | c |}
\hline
	\textbf{Cluster Indices} & \textbf{Sample\_Size} & \textbf{Z} & \textbf{J} & \textbf{H} & \textbf{K} & \textbf{SFR} & \textbf{U-G} \\ \hline
	\textbf{\emph{K1}} & 1616 & -21.93775(0.01694) & -21.66014(0.01993) & -22.38373(0.02041) & -22.82018(0.02079) & 1.0638(0.02713) & 1.27181(0.00525) \\ \hline
	\textbf{\emph{K2}} & 5257 & -21.7555(0.01212) & -21.55149(0.01157) & -22.28546(0.01128) & -22.71686(0.01084) & 0.29182(0.00681) & 1.51554(0.0054) \\ \hline
	\textbf{\emph{K3}} & 4103 & -21.49019(0.01297) & -21.25062(0.01396) & -22.01234(0.01381) & -22.4539(0.01368) & 0.2581(0.00657) & 1.47697(0.00511) \\ \hline
	\textbf{\emph{K4}} & 126 & -22.73229(0.07426) & -22.70672(0.06646) & -23.40741(0.06816) & -23.76453(0.06688) & 0.05794(0.0218) & 1.92586(0.02941) \\ \hline
	\textbf{\emph{K5}} & 425 & -20.58465(0.04568) & -20.41109(0.04792) & -21.18188(0.04722) & -21.69543(0.04628) & 0.46424(0.02539) & 1.10996(0.01159) \\ \hline
	\textbf{\emph{K6}} & 17 & -17.93447(0.91007) & -21.39051(0.33567) & -22.2191(0.28551) & -22.70181(0.30081) & 1.74706(0.80901) & 0.65155(0.98487) \\ \hline
	\textbf{\emph{K7}} & 23 & -19.92013(0.12171) & -19.70275(0.17297) & -20.48132(0.18368) & -20.87053(0.17381) & 0.41304(0.07394) & 0.94851(0.02521) \\ \hline
	\textbf{\emph{K8}} & 3723 & -21.32314(0.01359) & -21.0286(0.0149) & -21.7904(0.01453) & -22.2578(0.01449) & 0.26242(0.00661) & 1.42762(0.00444) \\ \hline
	\textbf{\emph{K9}} & 1 & -18.31074(0) & -18.28723(0) & -18.47423(0) & -19.55423(0) & 0(0) & 7.87047(0) \\ \hline
	\textbf{\emph{K10}} & 185 & -20.93675(0.05492) & -20.51798(0.04727) & -21.33757(0.04803) & -21.83218(0.04165) & 0.21892(0.02679) & 1.40617(0.0191) \\ \hline
	\textbf{\emph{K11}} & 6365 & -22.24511(0.01161) & -21.97196(0.0123) & -22.70899(0.01216) & -23.1256(0.01199) & 0.2798(0.00683) & 1.61358(0.00361) \\ \hline
	\textbf{\emph{K12}} & 4248 & -21.50759(0.01276) & -21.25928(0.01321) & -22.02619(0.01295) & -22.46407(0.01271) & 0.27093(0.00671) & 1.46643(0.00416) \\ \hline \hline
	\textbf{Cluster Indices} & \textbf{Sample\_Size} & \textbf{Z} & \textbf{J} & \textbf{H} & \textbf{K} & \textbf{SFR} & \textbf{U-G} \\ \hline
	\textbf{\emph{G1}} & 432 & -20.73857(0.03356) & -20.44912(0.03313) & -21.35022(0.03301) & -21.87317(0.03276) & 0.16505(0.01129) & 1.40178(0.01126) \\ \hline
	\textbf{\emph{G2}} & 3088 & -21.73313(0.01438) & -21.43728(0.01589) & -22.16386(0.01593) & -22.60139(0.0161) & 0.72847(0.01466) & 1.30725(0.00388) \\ \hline
	\textbf{\emph{G3}} & 6803 & -21.26019(0.00889) & -20.92901(0.00985) & -21.72344(0.00988) & -22.20691(0.00984) & 0.26987(0.00437) & 1.38886(0.00289) \\ \hline
	\textbf{\emph{G4}} & 10697 & -22.1379(0.00796) & -21.92457(0.00789) & -22.63046(0.00805) & -23.02487(0.00805) & 0.24855(0.00469) & 1.61023(0.0024) \\ \hline
	\textbf{\emph{G5}} & 101 & -20.59199(0.24866) & -21.21329(0.11427) & -21.9583(0.11029) & -22.4513(0.1088) & 0.71089(0.17431) & 1.30818(0.22496) \\ \hline
	\textbf{\emph{G6}} & 759 & -21.42326(0.03195) & -21.46964(0.03088) & -22.2503(0.02953) & -22.71963(0.02979) & 0.10198(0.00863) & 1.75566(0.02819) \\ \hline
	\textbf{\emph{G7}} & 2004 & -21.27896(0.01708) & -21.01205(0.01676) & -21.81426(0.01646) & -22.27631(0.01605) & 0.19596(0.00756) & 1.48726(0.00633) \\ \hline
	\textbf{\emph{G8}} & 704 & -20.54253(0.03133) & -20.35409(0.03294) & -21.17807(0.03232) & -21.69468(0.03311) & 0.34261(0.01514) & 1.15943(0.00788) \\ \hline
	\textbf{\emph{G9}} & 811 & -22.49597(0.02391) & -22.26266(0.02501) & -23.00113(0.02572) & -23.44847(0.02589) & 0.79063(0.03991) & 1.50365(0.01025) \\ \hline
	\textbf{\emph{G10}} & 416 & -22.93372(0.03575) & -22.77189(0.02965) & -23.46636(0.03046) & -23.82615(0.03035) & 0.05288(0.00979) & 1.8952(0.01016) \\ \hline
	\textbf{\emph{G11}} & 270 & -21.42426(0.06471) & -21.05023(0.06567) & -21.80613(0.06565) & -22.23182(0.06423) & 0.90222(0.06998) & 1.16632(0.01662) \\ \hline
	\textbf{\emph{G12}} & 4 & -18.67572(1.60019) & -20.17417(1.14757) & -20.74217(1.22211) & -21.17692(1.101) & 0(0) & 3.25642(1.54529) \\ \hline

\end{longtable}
\end{landscape}

\begin{landscape}
\begin{longtable}[c]{ | c | c | c | c | c | c | c | c |}
\hline
	\textbf{Cluster Indices} & \textbf{Sample\_Size} & \textbf{G-R} & \textbf{R-I} & \textbf{I-Z} & \textbf{U-Z} & \textbf{J-H} & \textbf{H-K} \\ \hline
	\textbf{\emph{K1}} & 1616 & 0.61146(0.00341) & 0.35539(0.00166) & 0.26177(0.00171) & 2.50043(0.01098) & 0.72359(0.00568) & 0.43645(0.00606) \\ \hline
	\textbf{\emph{K2}} & 5257 & 0.68736(0.00201) & 0.37617(0.00095) & 0.28002(0.00103) & 2.8591(0.00793) & 0.73397(0.00451) & 0.4314(0.00466) \\ \hline
	\textbf{\emph{K3}} & 4103 & 0.67019(0.00235) & 0.37471(0.0018) & 0.27525(0.00161) & 2.79711(0.00861) & 0.76172(0.00562) & 0.44155(0.00577) \\ \hline
	\textbf{\emph{K4}} & 126 & 0.86367(0.00633) & 0.42165(0.00452) & 0.32938(0.00546) & 3.54056(0.03394) & 0.70069(0.01356) & 0.35712(0.01082) \\ \hline
	\textbf{\emph{K5}} & 425 & 0.41384(0.00715) & 0.26574(0.004) & 0.17749(0.00355) & 1.96702(0.02333) & 0.77079(0.01829) & 0.51355(0.01996) \\ \hline
	\textbf{\emph{K6}} & 17 & 3.22742(0.67742) & 0.15845(0.18426) & -3.24598(0.93807) & 0.79144(0.79693) & 0.82859(0.09596) & 0.48271(0.08432) \\ \hline
	\textbf{\emph{K7}} & 23 & 0.3189(0.02483) & 0.18001(0.01508) & 0.13755(0.01397) & 1.58497(0.05698) & 0.77857(0.1127) & 0.38922(0.08759) \\ \hline
	\textbf{\emph{K8}} & 3723 & 0.63801(0.00252) & 0.35506(0.00128) & 0.26178(0.00131) & 2.68247(0.00883) & 0.7618(0.00632) & 0.4674(0.00648) \\ \hline
	\textbf{\emph{K9}} & 1 & 0.60358(0) & 0.33287(0) & 0.12404(0) & 8.93096(0) & 0.187(0) & 1.08(0) \\ \hline
	\textbf{\emph{K10}} & 185 & 0.62065(0.01164) & 0.34741(0.00594) & 0.25872(0.00618) & 2.63294(0.03915) & 0.81959(0.03145) & 0.49461(0.03302) \\ \hline
	\textbf{\emph{K11}} & 6365 & 0.76131(0.00188) & 0.4041(0.0019) & 0.31391(0.00195) & 3.0929(0.00683) & 0.73703(0.00303) & 0.41661(0.00317) \\ \hline
	\textbf{\emph{K12}} & 4248 & 0.66292(0.00235) & 0.36695(0.00159) & 0.27412(0.00165) & 2.77042(0.00818) & 0.76692(0.00562) & 0.43787(0.00577) \\ \hline \hline
	\textbf{Cluster Indices} & \textbf{Sample\_Size} & \textbf{G-R} & \textbf{R-I} & \textbf{I-Z} & \textbf{U-Z} & \textbf{J-H} & \textbf{H-K} \\ \hline
	\textbf{\emph{G1}} & 432 & 0.62095(0.00689) & 0.35303(0.00391) & 0.26257(0.00424) & 2.63833(0.02381) & 0.90109(0.02149) & 0.52295(0.022) \\ \hline
	\textbf{\emph{G2}} & 3088 & 0.60308(0.00247) & 0.34852(0.00126) & 0.2535(0.00122) & 2.51236(0.00821) & 0.72658(0.00502) & 0.43753(0.00517) \\ \hline
	\textbf{\emph{G3}} & 6803 & 0.6123(0.00169) & 0.34915(0.00091) & 0.25396(0.00093) & 2.60427(0.00591) & 0.79443(0.00488) & 0.48347(0.00501) \\ \hline
	\textbf{\emph{G4}} & 10697 & 0.75581(0.00125) & 0.40164(0.00062) & 0.30777(0.00063) & 3.07546(0.00454) & 0.7059(0.00239) & 0.3944(0.00256) \\ \hline
	\textbf{\emph{G5}} & 101 & 1.14197(0.14978) & 0.27309(0.11667) & -0.33705(0.23185) & 2.38619(0.21734) & 0.74501(0.03715) & 0.493(0.03616) \\ \hline
	\textbf{\emph{G6}} & 759 & 0.77627(0.00509) & 0.42452(0.00292) & 0.31725(0.00322) & 3.2737(0.03104) & 0.78066(0.01374) & 0.46933(0.01393) \\ \hline
	\textbf{\emph{G7}} & 2004 & 0.67452(0.00332) & 0.37561(0.00178) & 0.28329(0.00189) & 2.82068(0.01205) & 0.80222(0.00945) & 0.46205(0.00947) \\ \hline
	\textbf{\emph{G8}} & 704 & 0.44103(0.00492) & 0.27228(0.00314) & 0.18547(0.00296) & 2.05821(0.01635) & 0.82398(0.01488) & 0.51661(0.01582) \\ \hline
	\textbf{\emph{G9}} & 811 & 0.74156(0.00487) & 0.39747(0.00772) & 0.31498(0.00776) & 2.95766(0.01799) & 0.73847(0.00631) & 0.44734(0.00622) \\ \hline
	\textbf{\emph{G10}} & 416 & 0.86792(0.00318) & 0.42896(0.00217) & 0.3327(0.0023) & 3.52479(0.01341) & 0.69446(0.00656) & 0.3598(0.00661) \\ \hline
	\textbf{\emph{G11}} & 270 & 0.50498(0.01069) & 0.30478(0.00538) & 0.21357(0.0048) & 2.18964(0.03459) & 0.7559(0.01862) & 0.42569(0.01913) \\ \hline
	\textbf{\emph{G12}} & 4 & 1.07286(0.30623) & 0.45926(0.08625) & 0.41868(0.18051) & 5.20721(1.29849) & 0.568(0.12765) & 0.43475(0.23419) \\ \hline

\end{longtable}
\end{landscape}

\section{Results}\label{section 4}
\

In the present work we have compiled a data set upto a red shift of $z$ < 0.06, consisting of unbarred, weak barred and strong barred spiral galaxies and including starburst, AGN and LINER. At first we have checked the data set for Gaussianity and found it to be non-Gaussian. We performed Independent Component Analysis for dimentionality reduction and used MSTD to find out the optimal number of Independent Components. Then we clustered the data set with respect to these Independent Components by K-means cluster analysis followed by finding the number of optimum groups. The optimum number of ICs is 14 and the number of co-herrent groups is 12 (Table 2). From Table 2, it is clear that group 9 (K9) is an outlier and groups 6 and 7 (K6 and K7) contain a very small number of galaxies. Hence we have not considered K9. Subsequently we performed Gaussian Mixture Model Based Clustering (GMMBC) for checking the robustness of the groups. We have found similar number of groups with one group containing only 4 members like K9. \textcolor{black}{Remaining 11 groups found by both the methods are more or less compatible with respect to membership as well as average values of the parameters (K1 $\rightarrow$ G2, K2 $\rightarrow$ G6, K3 $\rightarrow$ G7,  K4 $\rightarrow$ G10, K5 $\rightarrow$ G8, K6 $\rightarrow$ G9, K7 $\rightarrow$ G11, K8 $\rightarrow$ G1, K9 $\rightarrow$ G12, K10 $\rightarrow$ G3, K11 $\rightarrow$ G4, K12 $\rightarrow$ G5 ). Therefore we have discussed the physical properties of the groups found with respect to K-means cluster analysis.}

\begin{figure}[htb!]
\centering
\includegraphics[width=\textwidth]{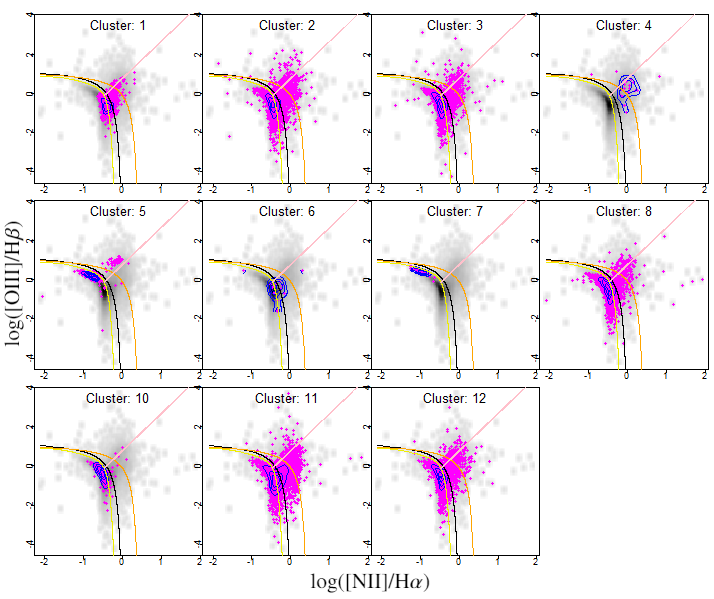}
\caption*{Fig. 6(a):  \textcolor{black}{log([OIII]/H${\beta}$) line ratios are plotted against log([NII]/H${\alpha}$) values for the whole sample (in grey) along with groups K1 - K12 (in magenta dots and blue iso contours) separately, except K9.} Curves are from \cite{stasinska2006semi}; \cite{kauffmann2003host}; and \cite{kewley2001theoretical} \textcolor{black}{(left to right respectively)}. The bottom left zone is for star-forming regions, the upper zone is for AGNs and Seyfert galaxies, and the bottom right zone is for LINERs.}
\end{figure}

\begin{figure}[htb!]
\centering
\includegraphics[width=\textwidth]{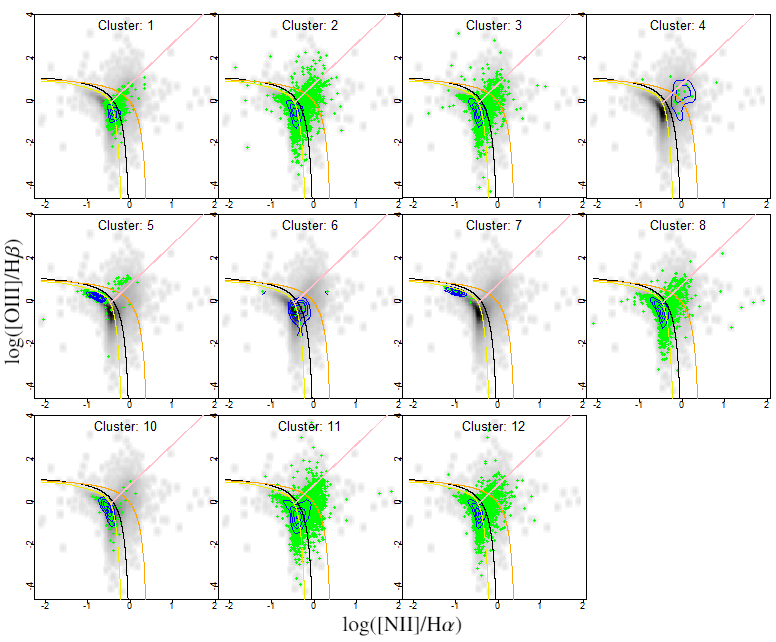}
\caption*{Fig. 6(b): Same as Fig. 6(a) but only for unbarred spiral galaxies (green dots).}
\end{figure}

\begin{figure}[htb!]
\centering
\includegraphics[width=\textwidth]{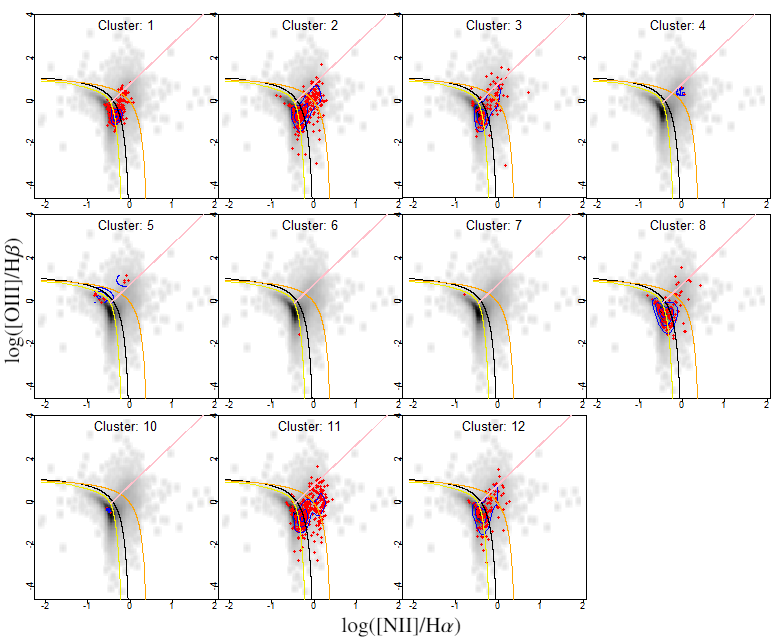}
\caption*{Fig. 6(c): Same as Fig. 6(a) but only for barred spiral galaxies (red dots).}
\end{figure}

\subsection{Properties of the ICs}
\

In the analysis of ICA, we  randomly selected set of ICs for which the variation in the multivariate set up is maximum with respect to Dunn index (Fig. 4). For the optimum set there are 14 ICs. ICs are actually linear combinations of several parameters (here 48 observable parameters) with various co-efficients. Few ICs are denoted by some features specific to a particular physical property of galaxies without any prior selection. At the same time each IC is not limited by the dominant feature. It also includes other with some lower weights and thus takes into account the complex interplay between observable parameters. Table 3 shows that among 14 ICs, 7 represent about five kind of properties:
\\
\begin{enumerate}
\item Metallicty (IC1,IC2,IC5),
\item Balmer absorption feature and low level ionisation (IC3),
\item Color (IC4),
\item Velocity dispersion (IC14),
\item Metallicity and high level ionisation (IC6, IC11).
\end{enumerate}
\

The remaining ICs are not dominated by any particular physical characteristic and the correlations with these ICs show negligibly small values. Thus even though we have got 14 ICs following \textcolor{black}{ IC1-IC14 (Subsections \ref{section 3.2.1} \& \ref{section 3.3.2} )} among those, 6, ICs have negligibly small effects on the total variation and 8 of them are the most significant ones.

\begin{figure}[htb!]
\centering
\includegraphics[width=\textwidth]{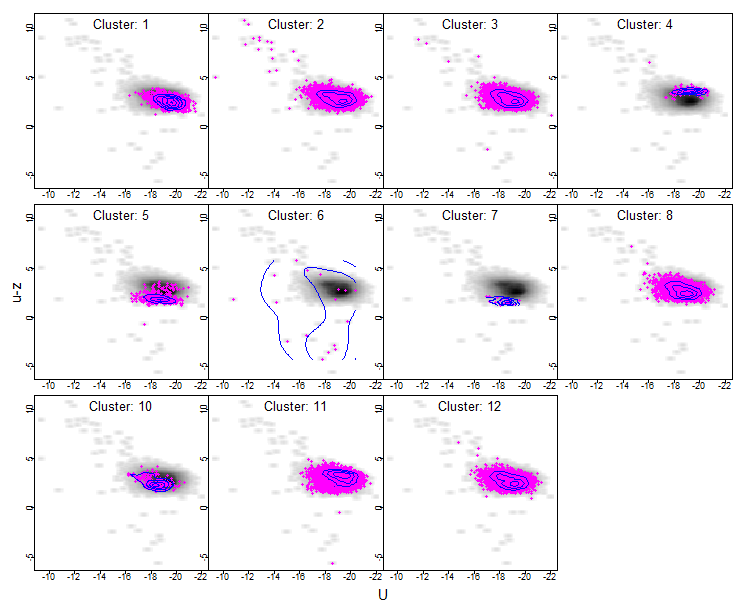}
\caption*{Fig. 7(a): Color–magnitude ($u-z$ vs. $U$) diagrams of the whole sample and for each of the groups K1 - K12 except K9.}
\end{figure}

\begin{figure}[htb!]
\centering
\includegraphics[width=\textwidth]{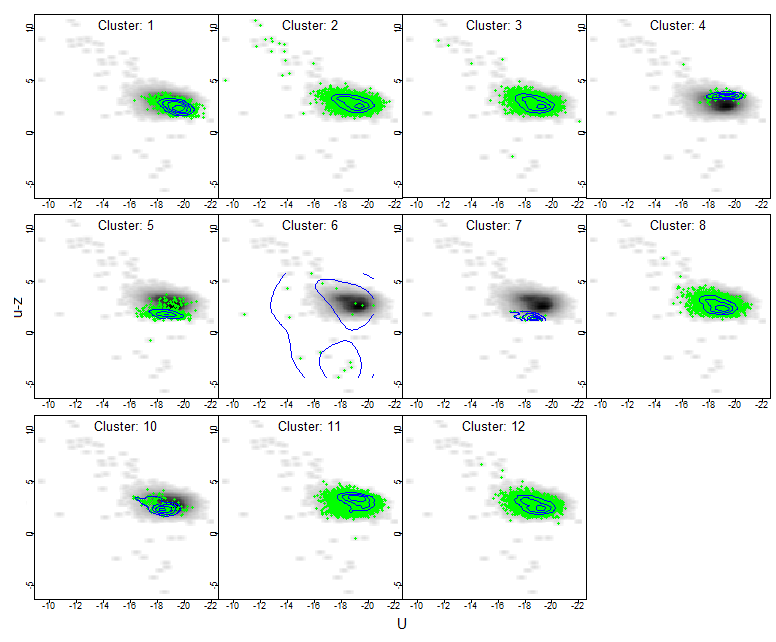}
\caption*{Fig. 7(b): Color–magnitude ($u-z$ vs. $U$) diagrams of Unbarred galaxies and for each of the groups K1 - K12 except K9.}
\end{figure}

\begin{figure}[htb!]
\centering
\includegraphics[width=\textwidth]{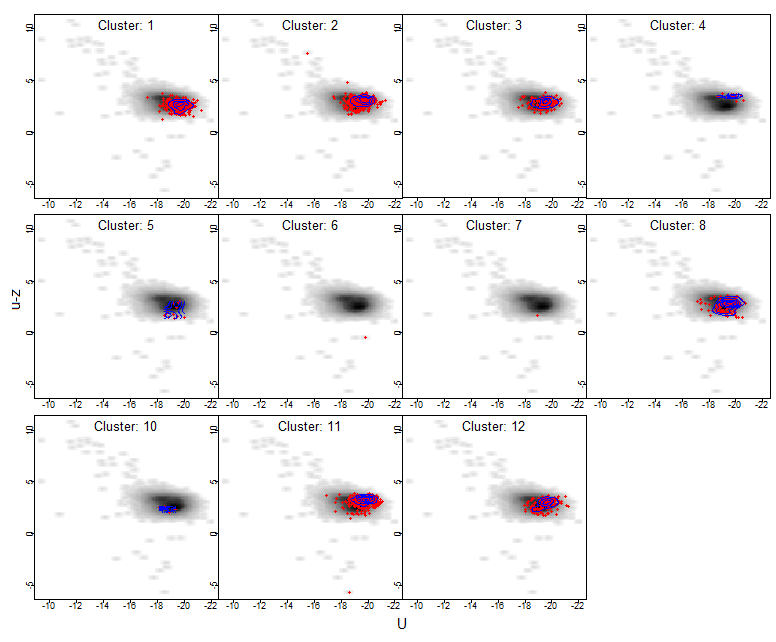}
\caption*{Fig. 7(c): Color–magnitude ($u-z$ vs. $U$) diagrams of Barred galaxies and for each of the groups K1 - K12 except K9.}
\end{figure}

\section*{Table 3}

\noindent(Observed parameters with highest correlation coefficients with significant ICs .)

{\renewcommand\arraystretch{1.5}
\noindent\begin{longtable}[c]{|l|l|}
\hline
\textbf{\emph{IC}}  & \textbf{Influential observed parameters} \\
\hline
\endfirsthead
$IC1$  & Lick\_Fe 5406 (0.54) \\
\hline
$IC2$ &    Lick\_Ca 4455 (-0.30) \\
\hline
$IC3$ & Lick\_H$_\beta$ (0.62), EW (Nii 6584) (0.65), EW (H$_\alpha$) (0.57), EW (H$_\beta$) (0.65), EW (H$_\gamma$) (0.65), EW (H$_\delta$) (0.65)  \\
\hline
$IC4$ & i-z (0.30) \\
\hline
$IC5$ & Lick\_Nad (0.76), Lick\_Ca 4227 (0.70), Lick\_g 4300 (0.76), Lick\_Fe 4383 (0.74), Lick\_Fe 4531 (0.71), \\
 & Lick\_C 4668 (0.79), Lick\_Mgb (0.77), Lick\_Fe 5406 (0.72), $D_n$(4000) (0.74), g-r (0.76), r-i (0.74),\\ & u-z (0.76)\\
\hline
$IC6$ & Lick\_Fe 5709 (-0.51)\\
\hline
$IC11$ & EW (Oiii 5007) (-0.64)\\
\hline
$IC14$  & $\sigma$-forbidden (0.38) \\
\hline
\end{longtable}}

\subsection{Properties of the galaxies in the groups.}
\

There are 11 effective groups (K1 - K8, K10 - K12), as a result of Cluster analysis. Among these K2, K4, K11 \textcolor{black}{(viz. Table 2)} consist of oldest galaxies \textcolor{black}{with respect to average ages and $D_n$(4000) values (though average age of K1 is higher than K2 but the average $D_n$(4000) values are just the opposite and since it is an observed parameter hence is more reliable one than the estimated value. Also K1 group of galaxies are similar to the groups of galaxies in the medium age range  with respect to other physical properties e.g. colour or metallicity etc.)}. K1, K3 ,K8, K10, K12 consist of galaxies of medium age, K6, K7 are the youngest groups of galaxies and K5 consists of unbarred galaxies. In each group we have classified 3 subgroups as strong barred (where the size of the bars are at least 30\% of their host galaxy size), weak barred (where the size of the bars is smaller than 30\% of the size of the host galaxy) and unbarred galaxies. It is clear from Table 2  that galaxies in groups
\textcolor{black}{ K2, K4 and K11 have galaxies of median ages, and \textcolor{black}{$D_n$(4000)} values which increase as, K2 < K11 < K4 and the median bar lengths of these groups also increase as K2 < K11 < K4.} The metallicities are higher in these galaxies. On the contrary \textcolor{black}{for galaxies in the groups K1, K3, K8, K10, K12 the median ages and $D_n$(4000) values vary more or less in the medium range ($\sim$ median age, 1.4 Gyr - 1.6 Gyr) with their bar lengths smaller compared to the oldest groups of galaxies.} Finally K6, K7 are the youngest groups of unbarred galaxies with minimum sample size and K5 is the youngest group (Median age $\sim$ 1.0 Gyr) containing barred and unbarred galaxies.

\begin{figure}[htb!]
\centering
\includegraphics[width=\textwidth]{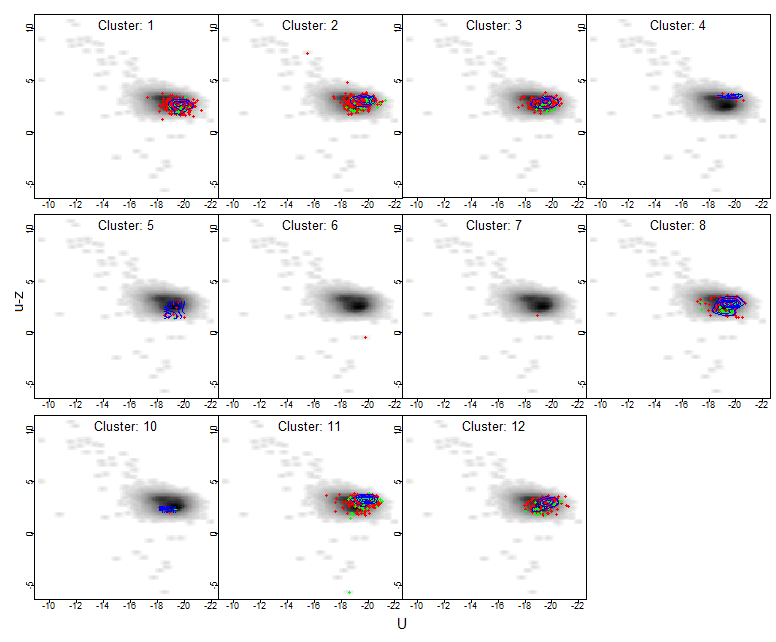}
\caption*{Fig. 8(a): Color–magnitude ($u-z$ vs. $U$) diagrams of the Barred galaxies and for each of the groups K1 - K12 except K9, where the red dots indicate the strong barred galaxies and the green dots indicate the weak barred galaxies.}
\end{figure}

\begin{figure}[htb!]
\centering
\includegraphics[width=\textwidth]{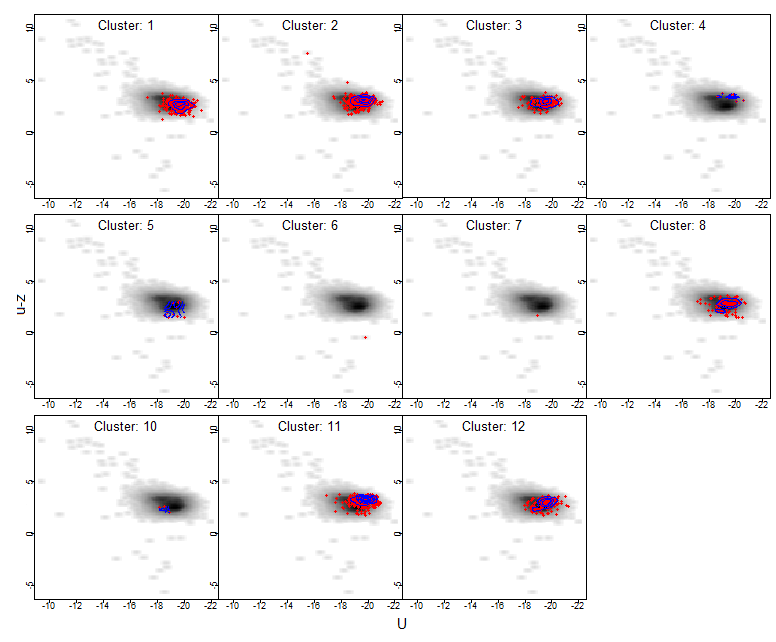}
\caption*{Fig. 8(b): Color–magnitude ($u-z$ vs. $U$) diagrams of the strong barred galaxies and for each of the groups K1 - K12 except K9.}
\end{figure}

\begin{figure}[htb!]
\centering
\includegraphics[width=\textwidth]{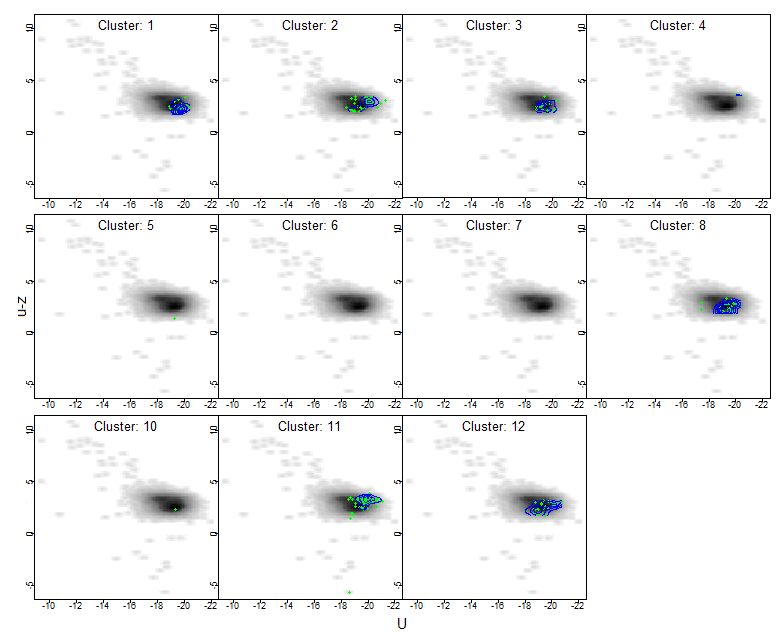}
\caption*{Fig. 8(c): Color–magnitude ($u-z$ vs. $U$) diagrams of the weak barred galaxies and for each of the groups K1 - K12 except K9.}
\end{figure}

\subsection{Emission line diagnostics and CM diagrams}
\

Emission line ratios have been recommended by various authors (\cite{baldwin1981classification}; \cite{veilleux1987spectral}; \cite{kauffmann2003host}) for qualitative classification of galaxies. According to the above scheme, scatter diagram of two emission line ratios (log(NII/H$_\alpha$) and log(OIII/H$_\beta$)) are classified by equations of curves separating the different classes of starburst galaxies, AGN and LINERs (Fig. 6). It is clear from Fig. 6 that K2, K4 and K11 contain all types of galaxies and the starburst galaxies have a wide range of ionization (log(OIII/H$_\beta$) $\sim$ -4 to 0.5). K1, K3, K8, K10 and K12 groups are primarily dominated by starburst with few AGN and LINERs, wheras K6, K7, K5 are populated by starburst galaxies.\\

These observations are more or less consistent with the groups. As K2, K4 and K11 are the oldest groups of galaxies and they contain both unbarred and barred galaxies. Star formation still occurs in some unbarred galaxies accompanied by AGN and LINERs which are oldest in ages. In galaxies of medium ages in the groups K1, K3, K8, K10 and K12, they are dominated by star forming galaxies rather than AGN or LINERs where star formation is being affected by the presence of bars (\cite{athanassoula1983formation}; \cite{buta1996galactic}). This is also reflected in  Fig. 7 where for the oldest groups (K2, K4, K11) the contours peak to redder color, medium aged groups (K1, K3, K8, K10, K12) concentrate around bluer zone and in particular in Fig. 8 where for the strong barred galaxies the contours comparatively peak at redder zone contrary to weak barred galaxies which peak at bluer zone in the color magnitude (CM) diagrams.\\

The color magnitude diagrams in ( $g-r$ , $R$ ) and color-color diagrams in ( $g-r$ , $u-r$ ) (Figs. 9(a) and 9(b)) for all galaxies (strong barred, weak barred and unbarred) show similar features as in Fig. 8. One interesting feature in color-color diagram shows that in K2 and K11 the weak barred galaxies occupy the same redder region as the strong barred galaxies. Since these galaxies fall in the oldest groups of galaxies, it indicates that these weak barred galaxies were previously strong barred galaxies but the bars are getting dissolved, supporting a recurrent phenomenon of bar formation. In the medium aged groups of galaxies the effect is not very pronounced due to non-availability of data on weak barred galaxies.

\begin{figure}[htb!]
\centering
\includegraphics[width=\textwidth]{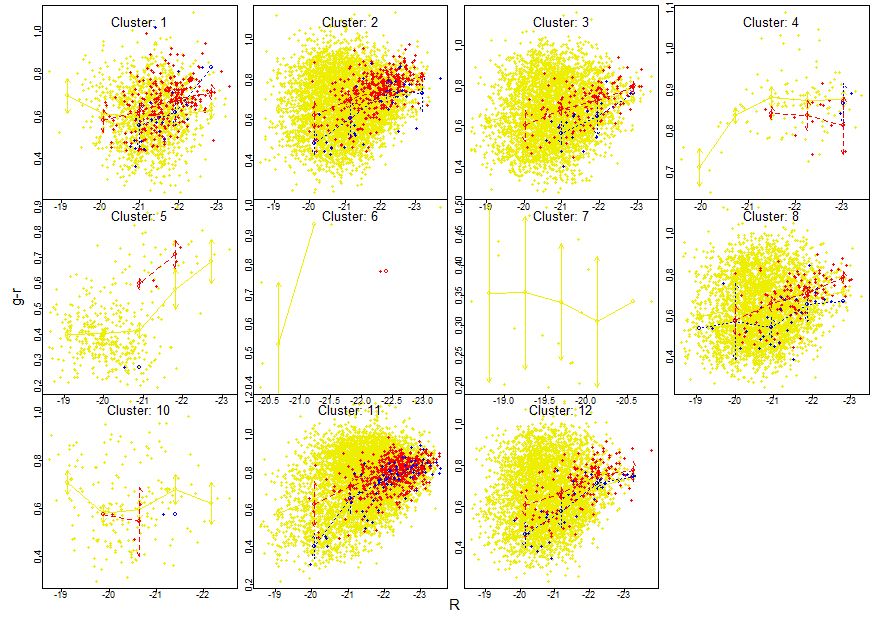}
\caption*{Fig. 9(a): Colour distributions $(g-r)$ are plotted against $R$ values for the groups K1 - K12 except K9 where the yellow dots indicate unbarred galaxies, red dots indicate strong barred galaxies and the blue dots indicate weak barred galaxies.}
\end{figure}

\begin{figure}[htb!]
\centering
\includegraphics[width=\textwidth]{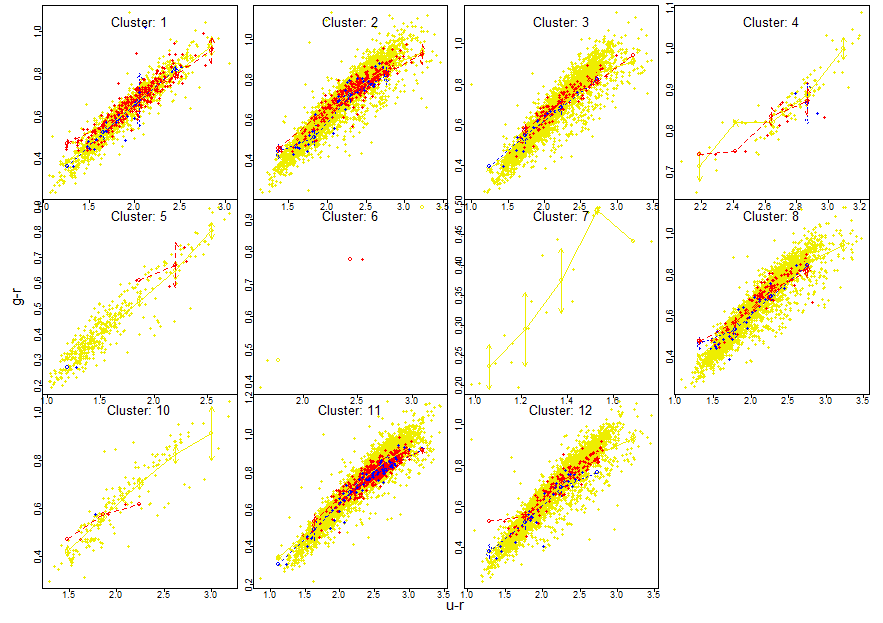}
\caption*{Fig. 9(b): Colour distributions $(g-r)$ are plotted against $(u-r)$ values for the groups K1 - K12 except K9 where the yellow dots indicate unbarred galaxies, red dots indicate strong barred galaxies and the blue dots indicate weak barred galaxies.}
\end{figure}

\subsection{Star formation efficiency}
\

Our aim is to assess the effect of bar on various properties of galaxies. Therefore we use the specific star formation rate parameter (\textcolor{black}{Star Formation Efficiency, hereafter SFE}) denoted by log($SFR/M_*$) \textcolor{black}{(viz. Table 1)} as a good measure for star formation activity in galaxies (\cite{brinchmann2004physical}). Also $D_n$(4000) (\cite{kauffmann2003host}; \cite{balogh1999differential}) has been used for the second estimate of age of stellar populations. Fig. 10 shows the SFE and $D_n$(4000) vs $log(M_*)$ for various groups K1 - K12. For the oldest groups i.e, in K2, K4, K11, data for K4 are not available.\\

For  K2, low mass weak barred galaxies have lower SFE compared to strong barred and unbarred galaxies ( $M_* \leq$ 10$^{10.5} M_\odot$ ). This might indicate that these galaxies were initially strong barred and their bars are getting  dissolved and \textcolor{black}{their SFE got} attenuated in the presence of \textcolor{black}{strong bar initially}. Thus they have low SFE (\cite{vera2016effect}). High mass weak barred galaxies ($M_*$ > 10$^{10.5}M_\odot$) have their bars growing from unbarred galaxies, which is quenching their star formation activity and they have again low SFE. Generally SFE is always lower in strong barred galaxies than unbarred galaxies. Various authors (\cite{masters2012galaxy}; \cite{ho1997search}; \cite{sheth2005secular}; \cite{ellison2011impact}; \cite{lee2012bars}) have suggested that the presence of bars funnel the gas into the central region of the galaxy. Subsequently the material turned into molecular gas which when becomes gravitationally unstable undergo fragmentation and trigger star formation activity. Therefore presence of bar could accelerate gas consumption ceasing formation of new stars in the outer region of the discs as they become redder. Thus in the star burst phase large amount of gas is transported towards the galactic central region for triggering star formation activity and in the post star burst phase the gas is consumed by circum nuclear star burst which shows low star formation rate (SFR) (\cite{jogee2005central}; \cite{sheth2005secular}).\\

In K4, SFE values are not available for barred galaxies. In K11, for high mass range ($10^{10}M_\odot$ > M > $10^{9}M_\odot$ ) strong barred and weak barred galaxies have similar SFE, which may be \textcolor{black}{as} low as $\sim 10^{-11}$ $yr^{-1}$). Few unbarred galaxies have very low SFE ( $\sim 10^{-11}$ $yr^{-1}$). \textcolor{black}{Since they are of oldest ages their SFEs are low.} Few weak barred galaxies in the high mass range have low SFE. This is indicative of the fact that these galaxies had strong bar initially but is getting dissolved. So, the SFE is low in these weak-barred galaxies. Some of this fact is consistent with the recurrent bar formation scenario in galaxies (\cite{kormendy2004secular}; \cite{bournaud2002gas}; \cite{berentzen2004regeneration}; \cite{gadotti2006lengths}; \textcolor{black}{\cite{katz2018gaia}; \cite{hilmi2020fluctuations}; \cite{de2019clocking}})\\

In the medium aged groups of galaxies K1, K3, K12, they have similar phenomenon of recurrent bar formation. e.g. in K1, K3 and K12 some features have been observed as in case of oldest groups.

\begin{figure}[htb!]
\centering
\includegraphics[width=\textwidth]{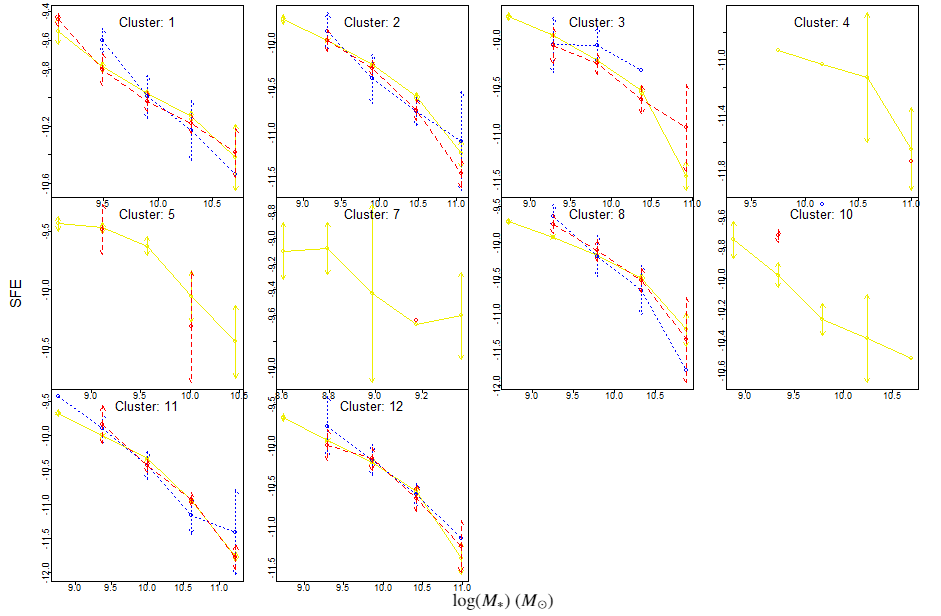}
\caption*{Fig. 10(a): SFE ($yr^-1$) values are plotted against log($M_*$) ($M_\odot$) for the groups K1 - K12 except K9 where the yellow line is for unbarred galaxies, red line is for strong barred galaxies and the blue line is for weak barred galaxies.}
\end{figure}

\begin{figure}[htb!]
\centering
\includegraphics[width=\textwidth]{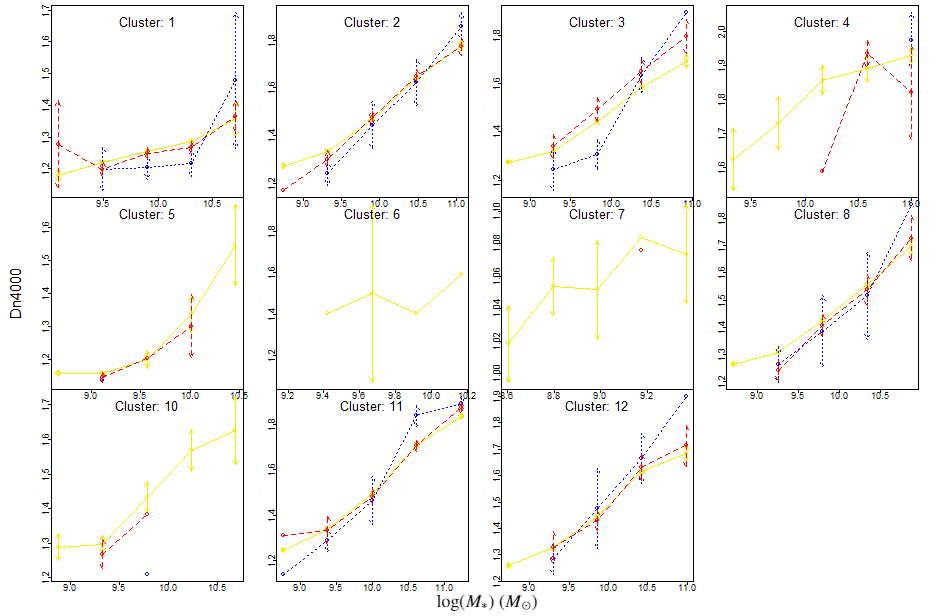}
\caption*{Fig. 10(b): D$_{n}$4000 values are plotted against log($M_*$)($M_\odot$) for the groups K1 - K12 except K9 where the yellow line is for unbarred galaxies, red line is for strong barred galaxies and the blue line is for weak barred galaxies.}
\end{figure}

\subsection{Mass-metallicity relation}
\

In Fig. 11 the mass-metallicity relation is shown for all the groups ( K1 - K12 \textcolor{black}{except K9)}. In the oldest groups of galaxies ( K2, K4, K11), barred galaxy mettalicity are always larger than unbarred ones. Metallicity increases with galaxy mass but the increase is almost constant after z > 0.02. For the same metallicity barred galaxy masses are higher than unbarred ones. In these groups the SFE decreases with increasing mass. This might be due to the fact that (\cite{ellison2007clues}; \cite{ellison2011impact}) metal enhancement without an accompanying increase in star formation activity may be due to a short lived phase of bar-triggered star formation in the past. Also the fall in the metallicity is rapid for strong barred galaxies rather than weak or unbarred galaxies. Weak barred galaxy curve crossed the strong bar curve in the high mass zone as well as in low mass range. This might be due to the fact that in the low mass range they can grow from strong barred ones as their bars are getting dissolved and from unbarred ones, in the high mass zone, \textcolor{black}{when} SFEs are low (viz. Fig. 10). For the \textcolor{black}{intermediate aged}  groups (K1, K3, K8, K10, K12) in most cases the data are not available from sdss but the more-or-less trend is similar as that of oldest groups. Moreover the crossing of weak bar curve with strong and unbarred ones indicates recurrent phenomenon of bar formation.

\begin{figure}[htb!]
\centering
\includegraphics[width=\textwidth]{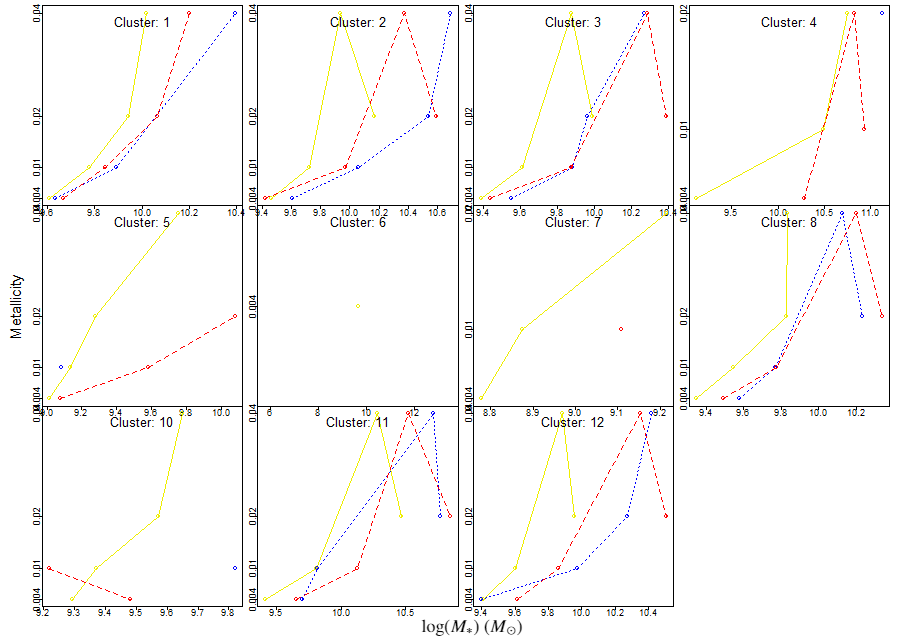}
\caption*{Fig. 11: Metallicity are plotted against log($M_*$)($M_\odot$) values for the groups K1 - K12 except K9 where the yellow line is for unbarred galaxies, red line is for strong barred galaxies and the blue line is for weak barred galaxies.}
\end{figure}

\section{Conclusion}\label{section 5}
\

The present work deals with a large data set of unbarred, strong barred and weak barred galaxies taken from sdss DR15 and cross-matched with zooniverse, for collecting bar properties of the barred galaxies. We have considered several significant observable parameters (e.g. Lick indices, Metallicity, SFR, Color Magnitudes etc.) (Table 1.) for performing the statistical analyses and the estimated parameters (e.g. age, SFE etc) along with observable ones are used for physical interpretation of the homogeneous groups. We have studied the influence of bars on the various properties of spiral galaxies. The following conclusions have been drawn from the above study:
\\
\begin{enumerate}
\item The entire data set is classified into 12 homogeneous groups among which group 9 is an outlier. Instead of parameters we have classified the data set with respect to 14, ICs which are linear combinations of various parameters with different weights. This is suitable for a non-Gaussian data set in a multivariate set up. We have found 12 groups by two independent methods, K-means cluster analysis (CA) and Gaussian Mixture Model Based Clustering (GMMBC) with respect to 14, ICs which establishes the robustness of the classification.
\item We have 14 Independent Components among which 8 components are found to be significant and they represent various influential physical galaxy properties like metallicity, ionisation, color, absorption features and velocity dispersion etc. Remaining 6 components do not \textcolor{black}{carry} much variation in the galaxy properties. With respect to these 14 components the data set has been classified into 12 homogeneous groups by K-means clustering and the robustness has been established by another widely used method GMMBC.
\item Among these 12 groups, four groups ( K2, K4, K11 ) fall in the oldest age ( $\sim$ 2.6 Gyr - 6.75 Gyr ) category and the groups ( K1, K3, K8, K12 ) fall in the medium range ( $\sim$ 1.68 Gyr - 1.8 Gyr ). One group is an outlier (K9) and the remaining two groups ( K6 - K7 ) are the youngest smallest groups of unbarred galaxies.
\item Oldest groups have longest bar lengths, highest metallicities and fall in the redder zone of color-magnitude diagram.
\item Galaxies of medium \textcolor{black}{age} range have shorter bar lengths and the groups are predominated by star burst galaxies, showing a kind of bluer zone.
\item In particular, \textcolor{black}{weak barred galaxies show indication of recurrent bar formation scenario} when the color-magnitude, color-color diagrams are studied thoroughly. This is consistent with the theoretical works suggested by various authors (\cite{kormendy2004secular}; \cite{gadotti2006lengths}).
\item It has been found that presence of bars may affect \textcolor{black}{the} SFE, metallicity, color magnitude and nature of galaxies. When the barred galaxies are oldest, they are redder, their bar lengths are longer and SFE are lower. Few weak barred galaxies, which are precursors of strong barred galaxies, have lower SFE and few weak barred galaxies which are precursors of unbarred galaxies of lower masses have higher SFE. This is the very new feature reflected in the present study and concludes that bar formation is not always one way phenomenon but may get dissolved in course of time in oldest and medium age ranged galaxies. 
\end{enumerate}

\section*{Acknowledgement}
\

\textcolor{black}{
The authors are grateful to the sdss Data Release 15 (sdss DR15) \footnote{\url{https://skyserver.sdss.org/dr15/en/tools/search/sql.aspx}} for giving the opportunity to retrieve the data and to work with it. One of the authors Prasenjit Banerjee acknowledges to a research fellowship provided by The University Grants Commission , with UGC Id. NOV2017-422665 \& UGC Reference No. 1141/(CSIR-UGC Net June 2017).
}

\bibliography{main}

\end{document}